\documentclass[usenatbib,useAMS]{mn2e}
\usepackage{epsfig,longtable,subfigure,graphicx,pdfpages}
\usepackage{epstopdf}
\usepackage{amsmath,amssymb}
\usepackage{float}
\usepackage[T1]{fontenc}
\usepackage{aecompl}

\title[Defocussed Transmission Spectroscopy: A potential detection of sodium in the atmosphere of WASP-12b] {Defocussed Transmission Spectroscopy: A potential detection of sodium in the atmosphere of WASP-12b}

\author[J.\,R.\ Burton et al.]  {J.\,R.\ Burton$^{1,}$$^2$\thanks{E-mail:
    jburton04@qub.ac.uk, jrburton@yorku.ca}, C.\,A.\ Watson$^1$,  P. Rodr\'{\i}guez-Gil$^{3,}$$^{4,}$$^5$,
  I. Skillen$^5$,  
   \newauthor S.\,P.\ Littlefair$^6$. S. Dhillon$^6$  and D.\ Pollacco$^7$\\
  $^1$ Astrophysics Research Centre, Queen's University Belfast, Belfast BT7 1NN, UK\\ 
$^2$ Department of Earth \& Space Science \& Engineering, York University, 4700 Keele St., Toronto, ON, M3J1P3, Canada\\
$^{3}$Instituto de Astrof\'\i sica de Canarias, V\'\i a L\'actea, s/n, La Laguna, E-38205, Santa Cruz de Tenerife, Spain\\
$^{4}$Departamento de Astrof\'\i sica, Universidad de La Laguna, La Laguna, E-38204, Santa Cruz de Tenerife, Spain\\
$^{5}$Isaac Newton Group of Telescopes, Apartado de correos 321, Santa Cruz de La Palma, E-38700, Spain\\
$^6$ Department of Physics and Astronomy, University
  of Sheffield, Sheffield S3 7RH, UK\\ 
  $^7$ Department of Physics, University of Warwick, Coventry 
  CV4 7AL, UK}

\date{\center{\Large Accepted for publication in the
    Monthly Notices of the Royal Astronomical Society \\
\vspace{.5cm} \today}}

\begin{document}
\maketitle

\begin{abstract} 
We report on a pilot study of a novel observing technique, defocussed transmission spectroscopy,
and its application to the study of exoplanet atmospheres using ground-based
platforms. 
Similar to defocussed photometry, defocussed transmission spectroscopy has an
added advantage over normal spectroscopy in that it reduces systematic errors due to
flat-fielding, PSF variations, slit-jaw imperfections and other effects associated with
ground-based observations.
For one of the planetary systems studied, WASP-12b, we report a tentative 
detection of additional Na absorption of 0.12$\pm$0.03[+0.03]\% during transit
using a 2\AA\ wavelength mask. After consideration of a systematic that occurs mid-transit,
it is likely that the true depth is actually closer to 0.15\%.
This is a similar level
of absorption reported in the atmosphere of HD209458b (0.135$\pm$0.017\%,
\citealt{snellen08}). 
Finally, we outline methods that will improve the technique during future 
observations, based on our findings from this pilot study.
\end{abstract}


\section{Introduction}\label{sec:intro}

The study of exoplanetary atmospheres has become one of the fastest developing
fields of astronomy in the last 10 years. Planetary atmosphere detection and 
characterisation have been 
achieved primarily using three methods -- multi-waveband transit photometry,
secondary eclipse photometry, and transmission spectroscopy. Multi-waveband transit photometry
involves primary transit observations in several wavelengths simultaneously. This probes 
to various depths in the exoplanet atmosphere, meaning the transit
lightcurve in each band can be fit in order to detect the difference in planetary radius
due to the transmission signal in the atmosphere
(e.g. \citealt{southworth12}, \citealt{haswell12}, \citealt{copperwheat13}). 
Secondary eclipse photometry
provides information on the thermal emission from the surface of the planet itself. Observations
of the secondary eclipse of hot-Jupiters in the visible and NIR 
(e.g. \citealt{gibson10}, \citealt{anderson10}, \citealt{burton12}) 
allow for the brightness
temperature to be calculated, an essential parameter for models of atmospheric simulations
and dynamics.
Transmission spectroscopy enabled the first detection of a hot-Jupiter exoplanet 
atmosphere (\citealt{charbonneau01}) where sodium was detected 
in the atmosphere of the transiting planet HD209458b. 
This is a technique where the in- and out-of-transit spectra
are compared and, depending on which absorbing species are present in the planetary atmosphere,
their presence can be determined by a deepening of the absorption features in the spectrum. 
Sodium is often studied due to the large relative signal it creates during transit relative to the surrounding continuum. In addition, the Na abundance can be probed to different depths, proving information on the temperature-pressure (T-P) profile of the atmosphere (e.g. \citealt{vidalmadjar11}, \citealt{huitson12}). 
The technique of transmission spectroscopy has since been carried out on a number
of hot-Jupiter exoplanets, both from space-based platforms 
(\citealt{sing08}, \citealt{knutson11}, \citealt{crouzet12}, \citealt{madhusudhan14}, \citealt{nikolov14}), 
and from the 
ground (\citealt{snellen08}, \citealt{crossfield11}, \citealt{bean11}, \citealt{mancini13}, \citealt{danielski14}).

Due to the contested 
nature of some of the detections of absorption features present in exoplanet atmospheres
(\citealt{swain08} vs. \citealt{gibson11}, \citealt{tinetti07} vs. \citealt{ehrenreich07}), it is imperative 
that any sources of uncertainty due to systematics are well-understood and characterised. 
Methods to mitigate such sources of uncertainty will allow for characterisation of smaller
exoplanets, in addition to planets orbiting fainter host stars.
Multiple observations of such systems will allow for
any claim of atmospheric detection to be verified by independent studies.
The characterisation of hot-Jupiters will also allow for any potential classification
systems which are based on atmospheric constituents (e.g. pM/pL -- \citealt{fortney08}, 
C/O ratio -- \citealt{madhusudhan12}) 
to be thoroughly 
tested, and provide much needed observational data points with which to test 
theories of exoplanet atmospheres (\citealt{marley13}). 
In regard to this, WASP-12b (\citealt{hebb09}) is an ideal candidate for carrying out
such tests. WASP-12b is a 1.4$M_{J}$ planet orbiting a G0 type star, on an orbital period
of 1.09 days. It is one of the hottest known hot-Jupiters, with an equilibrium temperature
of $\sim$2500K. Previous investigations into WASP-12b have revealed a high C/O
ratio (\citealt{madhusudhan11}, \citealt{stevenson14}), in addition to a transmission spectrum lacking in
TiO features (\citealt{sing13}). Such results are important aspects in testing classification systems
due to the predictions of atmospheric constituents which should be detectable in
hot-Jupiter atmospheres.
Recently, it has been discovered that WASP-12 is part of a triple system (\citealt{crossfield12}, 
\citealt{bergfors13}), meaning that
analysis must take into consideration blending with the two M-dwarf companions.

We propose a method of spectroscopic observation which aims to furnish the detection
of exoplanet atmospheres from ground-based systems, while reducing the impact of
systematics and errors arising due to flat-fielding and slit-loss corrections. In this paper, we report a pilot
study of a search for additional Na absorption in three exoplanet atmospheres.
Section \ref{sec:method} describes the method we adopted for our technique.
Section \ref{sec:obs} details the observation runs for the pilot study. Section \ref{sec:reduction}
describes the data reduction. 
Section \ref{sec:sys} describes the extensive data analysis we conducted, including
our search for systematics and the Na analysis in the atmosphere of WASP-12b.
Section \ref{DTS} discusses the technique in wider
context to future work and follow-up studies.
Section \ref{sec:future} contains our recommendations for future observations
utilising the technique, and finally, we provide some preliminary conclusions on the results 
from our pilot study.


\section{Method}\label{sec:method}

The typical precision required to detect an exoplanet atmosphere is of the order of 
$\sim$0.1\%.
Due to the systematics associated with ground-based observations, the technique of 
transmission spectroscopy provides a unique challenge. How does one minimise the
impact of systematics on observations whilst searching for an extremely small signal?
In a technique similar to defocussed photometry, we present the technique of `defocussed
transmission spectroscopy'. As with defocussed photometry, the aim of defocussed 
spectroscopy is to spread the 
light over a larger number of pixels on the CCD in order to reduce flat-fielding errors,
variations due to seeing, and other systematic sources of noise such as tracking errors. 
In contrast to high-resolution transmission spectroscopy, this method aims to maximise
the signal-to-noise (SNR) achieved from the ground. Since we are attempting to characterise
a strong individual spectral line, as opposed to the more subtle individual spectral features,
our aim is simply to maximise the SNR whilst using defocussing to minimise systematics. 
This makes the technique ideal for less-bright targets that high-resolution spectroscopy
has been unable to study in the past.
In order to detect the hot-Jupiter atmospheric features, particular lines in the spectrum are
selected (e.g. Na), which are then compared to the surrounding continuum during transit.
Any changes in the line depth due to the atmosphere annulus transiting the host star can 
then be detected by comparing the in- and out-of-transit line depths directly. This method
can be applied to any ground-based set-up for both low- and high-resolution spectroscopy.

\subsection{Simulations}\label{sec:sims}

\begin{figure}
\begin{center}
\includegraphics[scale=0.375,trim=5mm 5mm 1mm 1mm,clip=true]{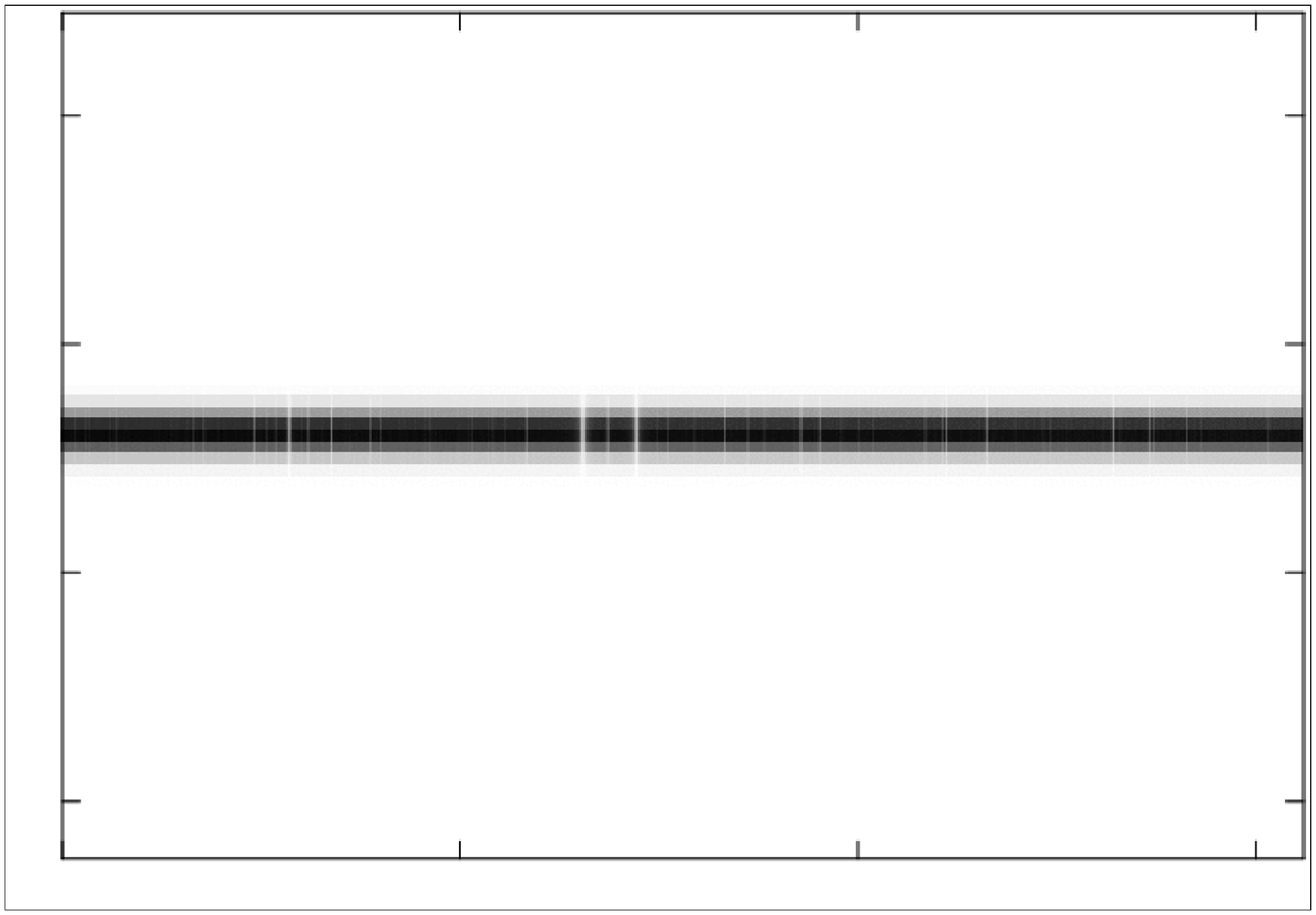}
\includegraphics[scale=0.375,trim=5mm 2mm 1mm 1mm,clip=true]{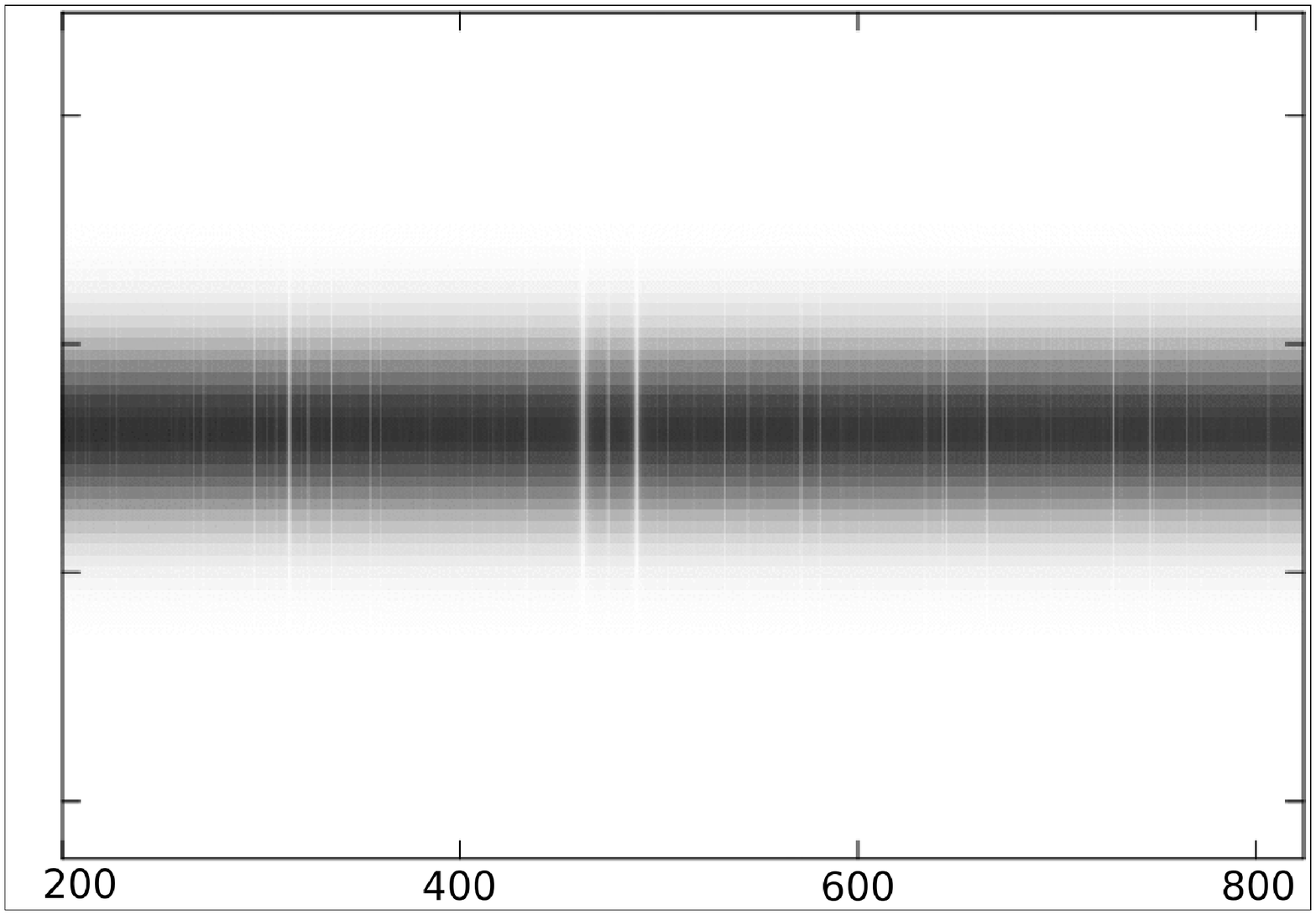}
\rule{15em}{0.25pt}
\caption[Defocussed transmission spectroscopy simulations]{Top: spectra in focus with a slit-width of  2.48". Bottom:
spectra defocussed to 2.48". Both simulations include the effects of variable seeing
conditions, slit jaw imperfections, flat-fielding errors and telescope tracking errors. The
two bright lines in the centre of each image correspond to the Na doublet. The x-axis displays the
pixels on the detector, and we have assumed a dispersion of 0.22\AA\ per pixel for the simulations.}
\label{fig:Spectra simulations}
\end{center}
\end{figure}

In order to investigate the effect of defocussing on systematic errors present in transmission
spectroscopy observations, we carried out a simulation.
Figure \ref{fig:Spectra simulations} shows a comparison between two simulated spectra, one with no defocussing and another defocussed to 2.48" with a slit width of 2.48", the widest slit possible
with an instrument like the Intermediate dispersion Spectrograph 
and Imaging System (ISIS) on the William Herschel Telescope (WHT), whilst maintaining adequate resolution (as obtained by \citealt{charbonneau02}). We aimed
to use the widest slit possible to reduce slit-losses due to differential 
refraction, as well as reducing the impact of any imperfections of the slit jaws that interfere with
the beam.

One major source of systematic noise during observations arises due to the motion of 
the spectrum over the CCD due to variations in seeing and drifts in position caused by
tracking errors and flexure (e.g. due to temperature variations). 
This not only causes the spectrum to broaden
and narrow in the spatial direction over consecutive exposures, but also moves the spectrum 
over different pixels. This can result in different spectra being incident on completely different
pixels over subsequent observations.
The aim of defocussing is to keep the bulk of the spectrum 
on the same pixels over the course of observations, and any motion that would
cause the spectra to move onto pixels with different responses will have a much
less significant effect. 
Furthermore, any change in the seeing will result in a smaller 
change compared to in-focus observations due to the spectrum spreading onto 
a smaller fraction of adjacent pixels. 

The model spectra in Figure \ref{fig:Spectra simulations} were generated using an 
observed solar spectrum convolved 
with a Gaussian in the spatial direction to simulate the effects of seeing and defocussing. 
The contribution of the broadened spectrum to 
each pixel in the simulation was then calculated, and then a pixel sensitivity map was used to 
mimic flat-fielding errors.
Slit-jaw imperfections were modelled
by simulating the impact of grains of dust up to 0.5$\mu$m, and the fraction of light blocked off at the specific spatial
location was multiplied by the intensity at the corresponding spectral position. The $\lambda^{-1/5}$ wavelength dependency of the FWHM of the seeing profile was also taken into account. Tracking errors
here used the typical errors found in WHT observations (up to 0.3 pixels), and
flat-fielding errors were at the 0.5\% level in order to account for additional sources
such as scattered light. Finally, seeing variations were modelled as having a FWHM that was randomly varied assuming a Gaussian distribution of 1.3"$\pm$0.3".
In comparison to the focused simulations, defocusing reduced the impact of systematic errors by an order of magnitude,
well within the precision required for an Na detection in the 
atmosphere of a hot-Jupiter exoplanet such as WASP-12b.

\subsection{Search for Na}\label{sec:na}

As mentioned earlier, Na is particularly targeted for 
ground-based studies  due to its expected strong signal during transit. In addition, Na is a very narrow feature
as opposed to the comparatively broad absorption bands of molecules 
such as CH$_4$, H$_2$O, TiO and VO. 
These molecular features are much more difficult to detect compared to the
sharp Na doublet absorption lines (located at 5890\AA\ and 5896\AA), even when using space-based instruments such as the HST/STIS. 
When testing potential classification systems -- such as the pM/pL system proposed by
\cite{fortney08} -- Na is an important distinguishing element between the two atmosphere
classes. Similar to M and L class stars, pM and pL class planets were proposed based
on their atmospheric constituents. pL class planets have a sufficiently low temperature to cause 
Ti and V to condense out lower down in the atmosphere. pM class planets, however, have a 
higher temperature, meaning gaseous TiO and VO can exist higher in the atmosphere. As
TiO and VO are very efficient absorbers of incident flux, this leads to pM class planets having 
anomalously bright daysides (due to a temperature inversion in the stratosphere caused by
TiO/VO absorption), as well as the Na absorption feature being suppressed. 
This means for pM class
planets, the level of Na absorption should be reduced, and hence should not be 
detectable in transmission spectroscopy observations. Na also has a history of being
detected in hot-Jupiter atmospheres, both from the ground and space (\citealt{charbonneau02},
\citealt{snellen08}, \citealt{sing08}, \citealt{redfield08}, \citealt{sing12} \citealt{nikolov14}).

We conducted a pilot study of this technique to detect Na in the atmospheres of three exoplanet
systems, WASP-12b, HD189733b and HD209458b. The latter two were chosen as these two bright 
objects ($m_{v}$$\sim$7.7) are the two most
well-characterised exoplanet atmospheres, with well-established Na detections for both
systems. If we were able to successfully detect additional Na absorption in these two systems,
that would provide an excellent demonstration of the technique. WASP-12 on the other hand
was chosen because of the large scale height of the planet ($\sim$1100km), given by:

\begin{equation}
H = \frac{kT}{mg},
\label{eqn:sh}
\end{equation}

\noindent where $H$ is the scale height of the atmosphere, $T$ is the temperature of the atmosphere, 
$m$ is the mean molecular mass and $g$ 
is the acceleration due to gravity. The fraction of stellar light which passes through the atmosphere ($F_S$) can be approximated by;

\begin{equation}
F_S = \frac{2{R_P}H}{{R_S}^2},
\label{eqn:fractionstell}
\end{equation}

\noindent where $R_P$ is the planetary radius and $R_S$ is the stellar radius (\citealt{pont09}).  For WASP-12b, the large scale height results in an 
expected transmission of stellar light which is $\sim$20\% greater than that of HD209458b
from equation \ref{eqn:fractionstell}. In addition, WASP-12b is firmly classified as a pM
planet in the \cite{fortney08} system, meaning the detection (or otherwise) of Na in the 
atmosphere of this hot-Jupiter provides a direct test of the classification system.

\section{Observations}\label{sec:obs}

Over all observations, we used the ISIS red arm with QuCAM mounted on the WHT with the R1200R grating. When observing in `Fast' mode, QuCAM has the ability to employ an electron multiplying CCD (EMCCD) amplifier.
The advantage of using an EMCCD is that one can
achieve excellent image quality with a short exposure time through the use of the EM gain register. 
This amplifies the signal in each pixel before reading out the image, and is used in both photometry and
spectroscopy for this very reason. The EM-gain also has the advantage of having a very low
effective read noise, something which benefits our specific observations since we are spreading
the light over a larger number of pixels. 
Another advantage of using QuCAM is the extremely short read-time, giving negligible dead-time between subsequent exposures. Not only does this increase the efficiency of the observations,
effectively boosting the S/N obtained, but also allows shorter exposures to be used with little loss
in on-target time. The ability to use shorter exposures is useful for two reasons; first it enables any relatively rapid variability due to systematics to be monitored and, second, it minimises the potential
for cosmic ray hits on the regions of interest that would potentially render individual spectra unreliable at the precisions required.

We decided against performing relative spectroscopy by placing another star on the slit as we
wished to prevent introducing potential systematics due to differential refraction. Instead, we chose to keep the slit orientated with the parallactic angle and measure the strength of the Na region compared to a relatively featureless region of nearby continuum, as previously implemented by \cite{snellen08}, for example. We opted to use the relatively featureless continuum region blueward of Na (from 5680-5885\AA)
as our comparison as \cite{sing08} found a source of additional absorption in the redward continuum of
HD209458b that, if present in our targets and used as the comparison region, would effectively reduce
the contrast of any additional Na absorption. The observational setup had a total wavelength coverage of 5680-5920\AA, with the Na doublet being positioned near the red-end of the spectral range for the
above reason. During these
observations, the dichroic was removed in order to reduce systematics due to ghosting,
an issue known to occur when using the dichroic in ISIS.
We also opted to use the red arm, again to reduce systematics, as this meant that the
light path did not have to be split off at an angle -- as would have been the case if we were to
have used the blue arm. The slit of 2.48" projects to 8 QuCam pixels, with the R1200R grating (taking
anamorphic magnification into account), and assuming a dispersion of 0.22\AA\ per pixel, this gives
1.76\AA\ for the 8 pixel projection.

WASP-12 was observed over 3 nights from 2nd-4th February 2010. Due to the short orbital period of this
exoplanet system, primary transits
of WASP-12b were visible on all nights. 
Unfortunately, the first night was lost due to
bad weather. The remaining 2 nights were clear, and transits of WASP-12b were obtained on
both nights, along with 1.3 hours of out-of-transit data on 3rd Feb and 1.2 hours on 4th Feb.
On the 4th Feb, we ended
observations during transit due to the position of the object on the sky. Upon
further investigation, it was discovered that the 4th Feb data were subject to
additional instrumental effects due to changing the CCD control head. Due to this,
in addition to having to end observations mid-transit, we opt to concentrate solely on the 3rd Feb data
in our search for Na absorption. During this night's observations, we were forced
to abandon a large series of data during ingress due to an issue with the telescope
control system. The `Fast' observing mode deletes any data taken during a run which
you are forced to abort, a significant factor in our decision to switch to `Slow' during the 
subsequent observations.
We note here that every time 
we moved off target (for example, to observe a telluric standard) we took an arc exposure.
This allows for the wavelength drift over the chip to be tracked throughout the night, and 
provides a wavelength calibration for all spectra. For these observations,
we defocussed the spectra to a FWHM of 12 pixels, corresponding to 2.4".

HD189733 and HD209458 were observed on the nights beginning 5th and 6th September 2010. 
Here, the
observing run was marred by bad weather and variable observing conditions throughout.
The most poorly affected time was during the transit of both objects. After looking at the
available data, it became clear that the in-transit spectra were too badly affected to get a
detection from. Nonetheless, the out-of-transit data was still used in order to test for
systematic effects and is presented. Throughout all observations, we aligned the slit with the
parallactic angle in order to reduce differential refraction, and, as during the previous
observing run, we defocussed to a FWHM of 12 pixels at the start of observations.
A summary of all observations is provided in table \ref{tab:tscand}.

\section{Data reduction}\label{sec:reduction}

The data were reduced using the \textsc{pamela} spectra reduction suite\footnote{http://deneb.astro.warwick.ac.uk/phsaap/software/}. Prior to input into \textsc{pamela}, 
all science images were de-biased and flat-fielded. Flat fields for the spectra were created
using a tungsten lamp directly in the path of the beam in order to generate the uniform source 
of light for each frame. A masterflat was created by median-stacking the individual frames
and subsequently removing the spectral response of the lamp using a polynomial fit.
An optimal extraction was used in
order to produce individual sky background fits for each exposure. This recipe was followed
for each exposure for every object. We note here that for some HD209458 observations, 
a nearby star appears present on the slit. In these cases, the offending object was masked out
during reduction.
The extracted data were then wavelength calibrated using an interpolation between the
two nearest arcs in time surrounding the spectra. Residual wavelength shifts were still
visible in the data after applying the arc calibration. In order to correct for this, we 
cross-correlated the data (using an out-of-transit high S/N spectra as a template), and then
removed the residual shifts given by the cross-correlation function. 
This is important given the method of measuring absorption we have chosen to adopt 
(see section \ref{sec:wasp12} for details), whereby we define a window around the Na doublet
and integrate the flux under the spectrum. If, however, our wavelength calibration is
incorrect, there is a potential to shift the Na line out of the window, or shift part of the continuum
into the window. This could potentially have the effect of causing the Na depth to vary
over subsequent spectra due to this motion.

\section{Data Analysis}\label{sec:sys}

The analysis of the spectra of our targets was carried out following the general approach undertaken by other transmission spectroscopy studies (e.g. \citealt{sing08}, \citealt{snellen08}). In these studies, the strength of the absorption was measured relative to a continuum window, and the in- and out-of-transit depths compared. For these purposes we integrated the flux under the spectra, covering pre-defined wavelength windows.
Any change in the flux ratio during transit may be
indicative of additional Na absorption due to the exoplanet atmosphere.
The data were analysed using the \textsc{molly} spectrum analysis software$^{1}$.

\subsection{HD189733 and HD209458}\label{sec:HD}

As mentioned in section \ref{sec:obs}, the in-transit data for HD189733b and HD209458b were
badly affected by variable cloud cover, poor seeing and adverse weather conditions.
Therefore, we opted instead to analyse the systematics associated with our observing
technique for the out-of-transit data obtained during these two nights instead.  

\begin{figure}
\includegraphics[width=\textwidth/2,height=4.75cm,trim=0mm 0mm 0mm 0mm,clip=true]{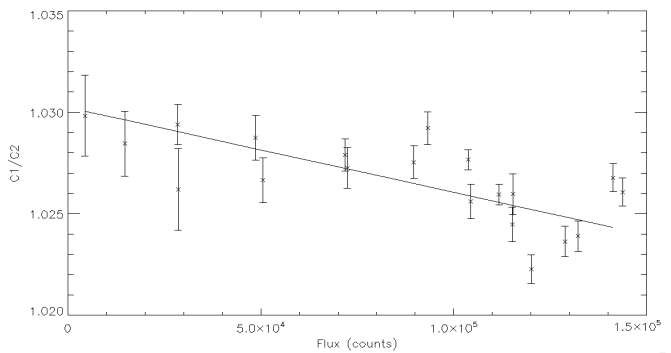}
\caption[C1/C2]{C$_{1}$/C$_{2}$ for the spectra of HD189733 and HD209458 for 10 different 
flux levels.
The trend here is linear, indicating that the spectra become more tilted towards the red end of the spectrum for lower flux
values. We note that WASP-12 is not included as the flux level remained relatively stable throughout
the observing run, with only a handful of spectra varying due to short-term transparency variations. This was due to better weather conditions and more steady transparency.}
\label{fig:c1c2errors}
\end{figure}

One of the main issues when carrying out transmission spectroscopy is the linearity
of the detector. If the instrumental response causes the deep Na features to appear
shallower due to the non-linearity of the detector, an absorption signal can easily be
mimicked. \cite{snellen08} investigated non-linearity effects for their HDS (High 
Dispersion Spectrograph) data 
of HD209458 by comparing the normalised continuum count level with the measured mean depth of 
59 identified absorption lines. This indicates a strong inverse correlation (i.e. for deeper absorption lines,
the higher the flux level in the continuum). In order to perform this analysis, a significant number of 
absorption features are required, something which we are lacking due to the relatively narrow wavelength
coverage we have for our observations. An excellent way in which to characterise the linearity is to
obtain flat-fields of different exposure lengths. This gives a uniform response to a range of count
levels, and can be used to construct a correction factor if needed. Unfortunately, we neglected to 
obtain flat-fields of varying exposure time for our data. We are able, however, to use the variable count
levels obtained during periods of poor weather to mimic this effect. We did this by first 
grouping spectra into differing flux levels, and performing an analysis for two different 
continuum windows with wavelength ranges 5720-5790{\AA} and 5790-5860{\AA} (hereafter 
referred to as C$_{1}$ and C$_{2}$, respectively). For a completely linear detector, the slope 
of the continuum should remain constant over all levels of flux. Figure \ref{fig:c1c2errors} shows the
continuum 
window C$_{1}$ against the ratio of C$_{1}$ to C$_{2}$. Note that Figure \ref{fig:c1c2errors} shows flux values 
for both HD189733 and HD209458 after renormalising them to the same scale. 
We note here that for both observing runs, we kept the count levels in the linear
regime
of QuCAM\footnote{http://www.ing.iac.es/Engineering/detectors/QUCAM2lin.jpg} for both `fast' and `slow' observing modes (see Table \ref{tab:tscand}). When using QuCAM in the fast observing mode, we
ensured to keep the counts well below 25K ADU, where the non-linearity issues are much worse.

\begin{figure}
\includegraphics[width=\textwidth/2,height=4.5cm,trim=0mm 2mm 2mm 0mm,clip=true]{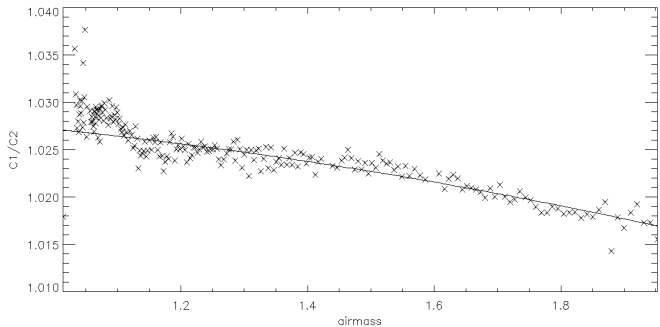}
\caption[C1/C2 vs.Airmass]{The ratio of C$_{1}$ to C$_{2}$ as a function of airmass for HD189733. 
The trend here indicates that as the airmass increases, the continuum ratio decreases.}
\label{fig:c1c1airmass}
\end{figure}

From Figure \ref{fig:c1c2errors}, it can be seen that at lower flux levels, the ratio is higher, meaning the spectra 
are essentially more sloped at lower overall flux values.  The question now becomes what is 
the cause of this difference in the ratios of the continuum flux levels. Is it solely due to a non-linearity 
of the detector, or could there be some other underlying cause? Upon further investigation, we 
determined a possible cause could be atmospheric extinction -- the effect of losses at a specific
wavelength due to absorption/scattering by the Earth's atmosphere. This effect would manifest 
as a correlation of the ratio of C$_{1}$ to C$_{2}$ with airmass. 
Figure \ref{fig:c1c1airmass} shows a linear correlation between the ratio and airmass, along with 
the trend modelled as a least-squares fit. This trend appears for all three objects, indicating
that this is a systematic inherent to the observations as a whole. 

From Figure \ref{fig:c1c1airmass}, it can be seen that the majority of the points at low airmass show a 
high C1/C2 flux
ratio. These are the points which were most affected by low count rates due to cloud coverage. In 
order to avoid these points contaminating the least squares fit (and hence causing a bias in the 
trend, favoured towards the nosier points), we used a Gaussian weighting factor, $W$, described by;

\begin{equation}
W = \frac{1}{{Y_i}^2},
\label{eqn:weighting}
\end{equation}

\noindent where $Y_{i}$ are the errors associated with the $Y$ values (in our case, the ratio of C1 to C2).
This trend was then removed from all the data sets using a weighted polynomial extraction.

After this term was removed, we searched the corrected continuum for any correlations
with parameters such as spatial position of the spectra on the CCD, FWHM, sky background, temperature
of the chip and peak continuum count level. No correlations could be found for the out-of-transit data not affected by poor weather conditions. However, in order to fully characterise
these systematics for this observing method, a data set observed under near-photometric conditions would be desirable in order to check for any additional correlations that may be present at a low level not identified in our current limited out-of-transit data set. 
We used identification and analysis of these systematics in these data sets to provide similar corrections 
to WASP-12 described in the next section.

\subsection{WASP-12}\label{sec:wasp12}

\begin{figure}
\begin{center}
\subfigure[]{%
            \label{fig:a}
            \includegraphics[scale=0.26,trim=0mm 0mm 0mm 0mm,clip=true]{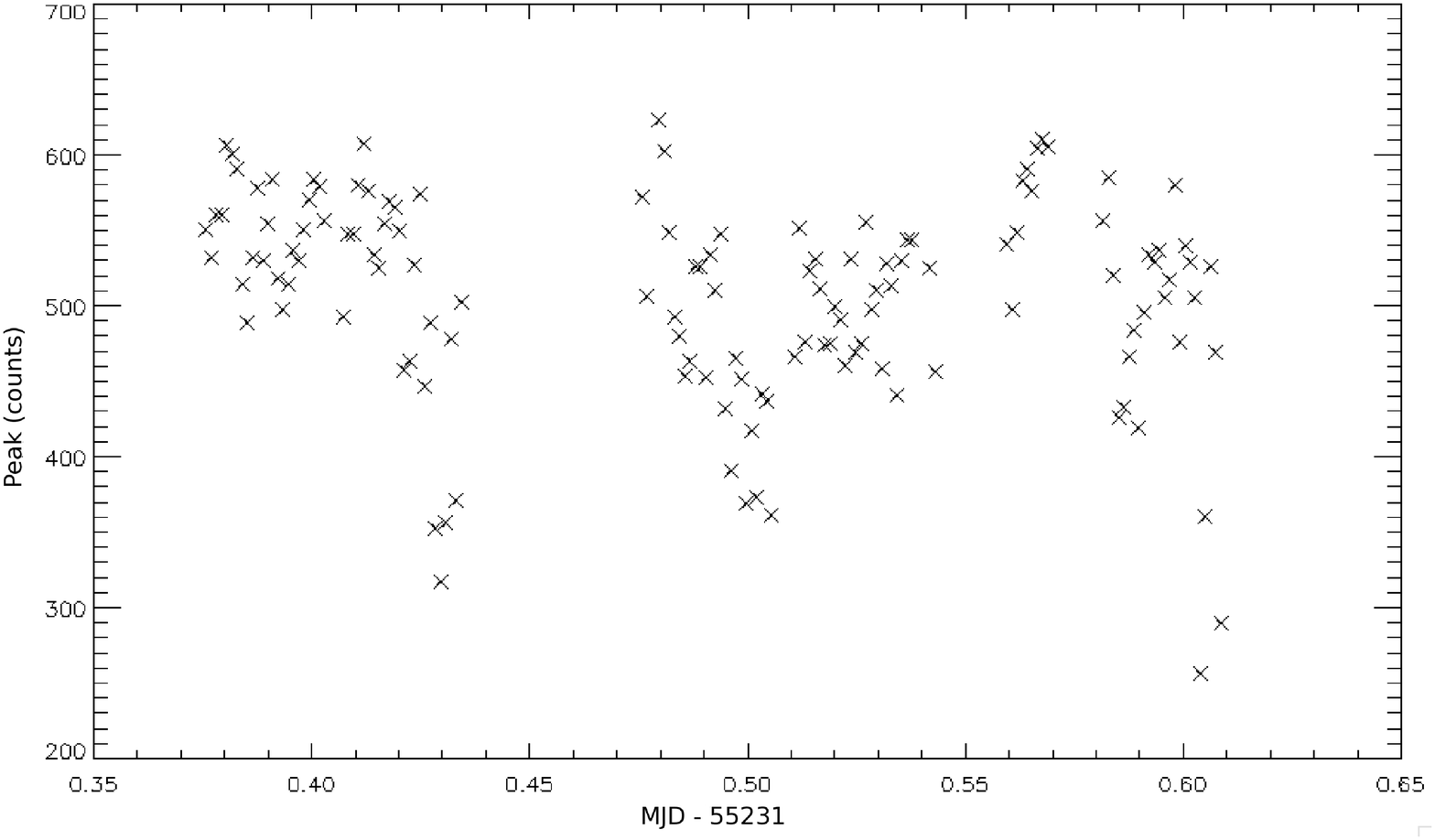}
        }%
        \subfigure[]{%
            \label{fig:b}
            \includegraphics[scale=0.26,trim=0mm 0mm 0mm 0mm,clip=true]{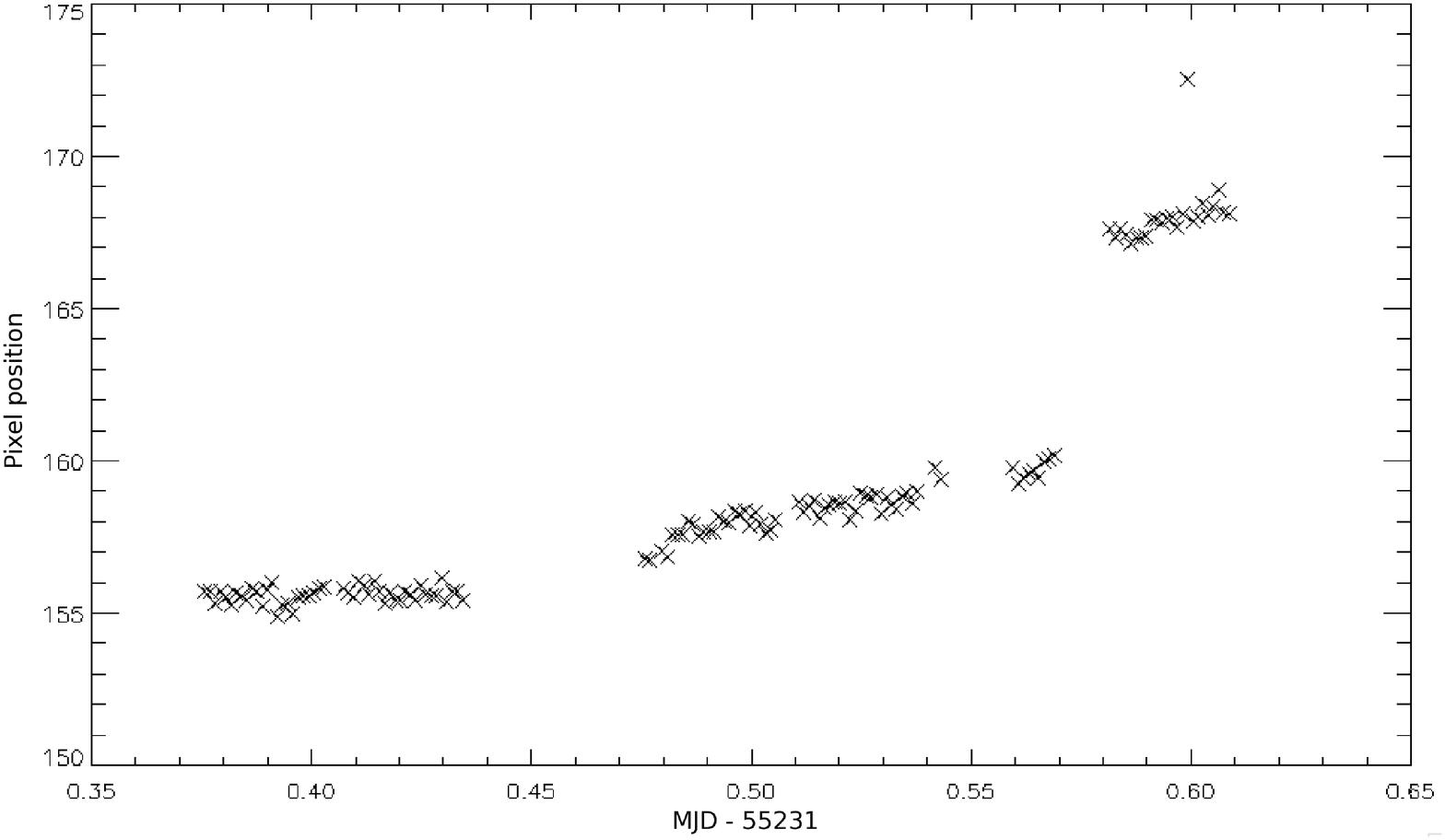}
        }\\ %
        \subfigure[]{%
            \label{fig:c}
            \includegraphics[scale=0.265,trim=4mm 0mm 0mm 0mm,clip=true]{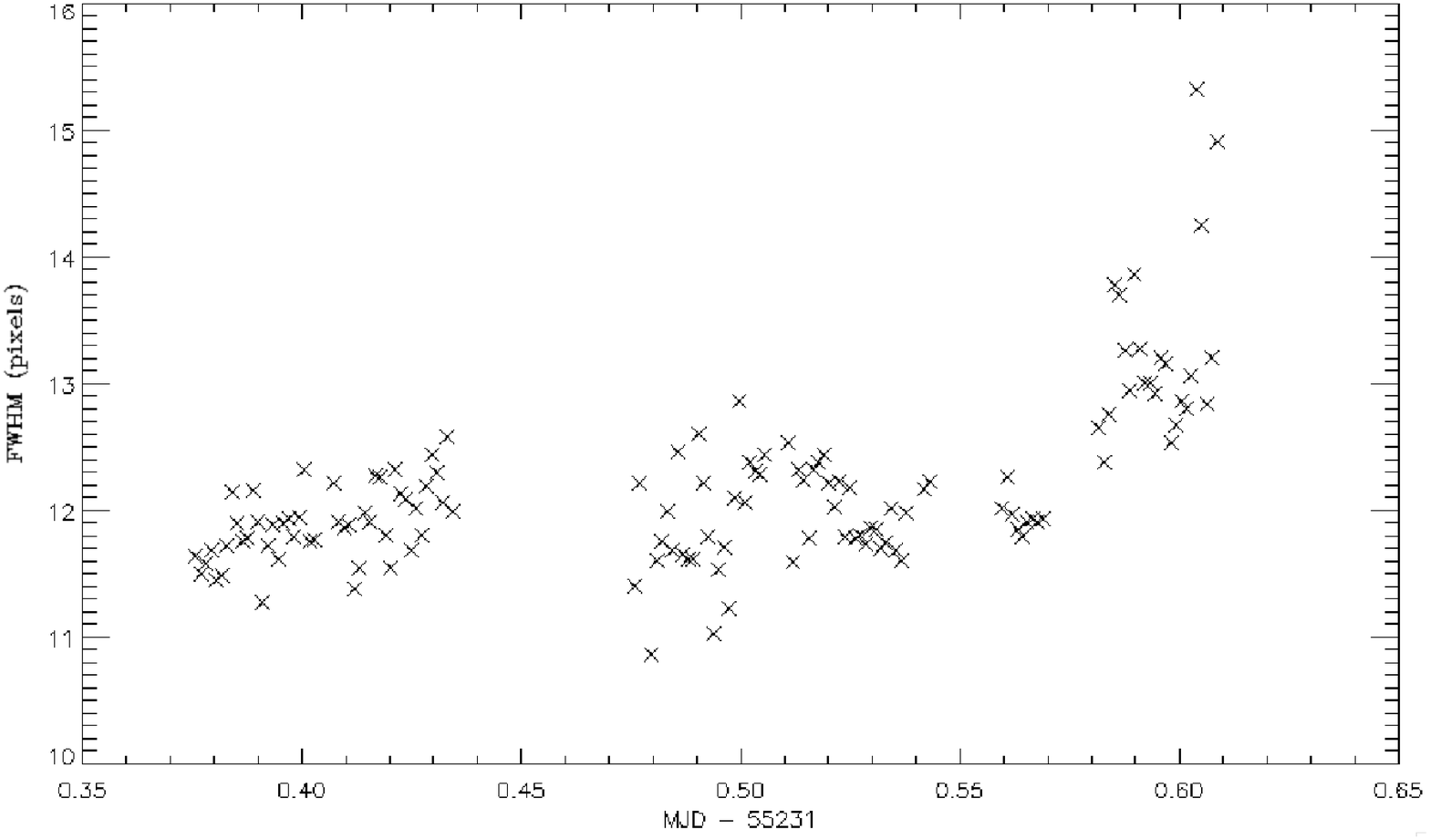}
        }%
              \subfigure[]{%
            \label{fig:d}
            \includegraphics[scale=0.2475,trim=-3mm 0mm 0mm -1mm,clip=true]{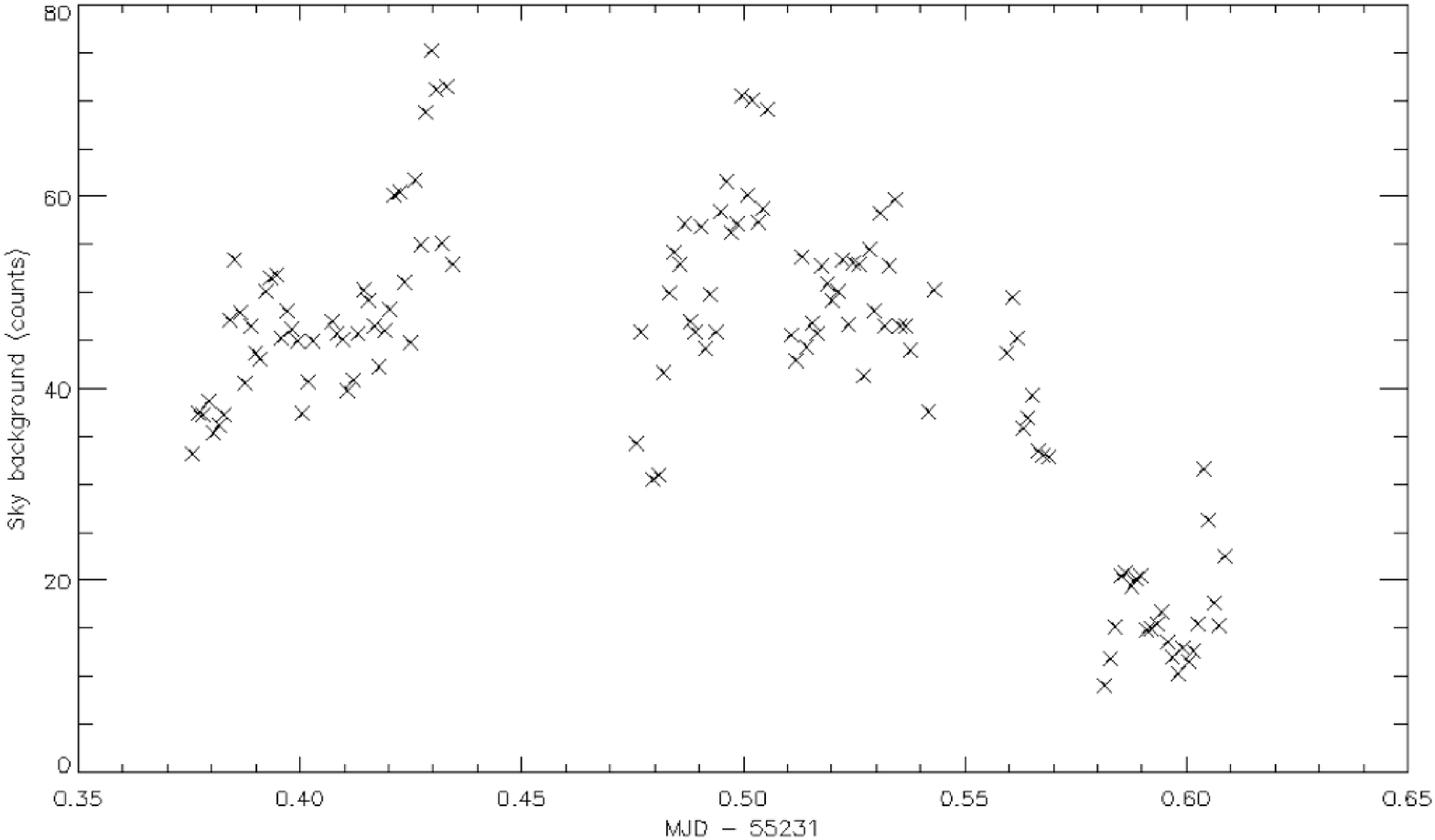}
        }\\%
        \caption{WASP-12 analysis -- MJD vs. (a) Continuum flux. (b) Position of spectra on the CCD. (c) FWHM (d) Sky.}
\label{fig:wasp12}
\end{center}
\end{figure}

In order to search for additional Na absorption in the WASP-12 
data, we performed an analysis with \textsc{molly} to integrate the flux within 
predefined wavelength windows for two different 
regions -- the Na doublet (5885-5900{\AA}) and the blueward 
comparison region of the continuum (5720-5860{\AA}) -- after we 
had performed the correlation analysis on this data in the same 
manner as described in section \ref{sec:HD} (see Figure \ref{fig:wasp12}). This matches the 
continuum windows used in the systematic analysis for HD189733 
and HD209458. In order to attempt to detect a deepening of the Na 
lines relative to the surrounding continuum, we adopted a similar 
approach to \cite{snellen08}. First, we normalised the spectra, and 
then performed the flux analysis using masks to integrate under 
different parts of the spectra. The final step, once all necessary correlations 
had been removed (as detailed in sections~\ref{sec:HD}), was to measure the 
ratio of the integrated flux in the two windows. This flux ratio should be constant 
(within error bars) throughout the night if we assume no Na absorption 
is taking place. However, if the lines do deepen due to the influence 
of the atmosphere of WASP-12b during transit, the integrated flux ratio 
will be different when comparing the in- and out-of-transit flux ratios.
During our observations, we were able to achieve a signal-to-noise (SNR)
of 200 per spectrum.

\begin{figure}
\includegraphics[scale=0.3,trim=0mm 0mm 0mm 0mm,clip=true]{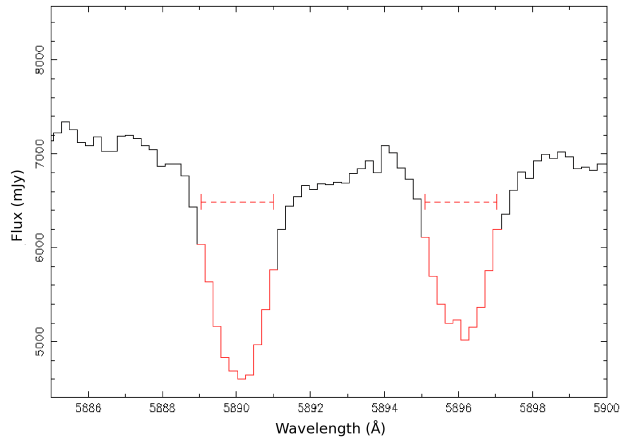}
\caption[Narrow mask]{The mask used in the search for Na absorption. The regions
used in the search are highlighted in red, and are 2{\AA} across.}
\label{fig:narrowmask}
\end{figure}

\cite{snellen08} report that the detection of Na in the atmosphere of 
HD208458b was significantly stronger when a narrow window was 
used. They used three different window sizes termed `narrow', 
`medium' and `wide', that were positioned over each line in the Na 
doublet, with widths 0.75, 1.5 and 3.0 {\AA}, respectively. The 
relative photometric dimming of the Na feature to the surrounding continuum
were deeper by 0.056$\pm$0.007\% (wide), 0.070$\pm$0.011\% (medium) and 
0.135$\pm$0.017\% (narrow). In order to determine if Na could be 
detected in our WASP-12 data, we opted to use a single window which 
would maximise the contrast of the Na region relative to the continuum. 
A number of different window sizes were trialled during this analysis, and
a 2.0\AA\ window over each component of the doublet was determined to 
be wide enough so as to avoid any 
issues such as clipping the Na line, but sufficiently narrow as to allow 
the contrast to be maximised. While this is wider than the `narrow' and 
`medium' windows used by \cite{snellen08}, this was necessary due to 
the lower resolution of the ISIS data. 
Figure \ref{fig:narrowmask} shows 
the masks used to analyse the Na doublet in the WASP-12 spectra, at 
5889 - 5891 and 5894 - 5896{\AA}. 

\begin{figure}
\subfigure[]{%
            \label{fig:a}
            \includegraphics[scale=0.31,,trim=0mm 0mm 0mm 12mm,clip=true]{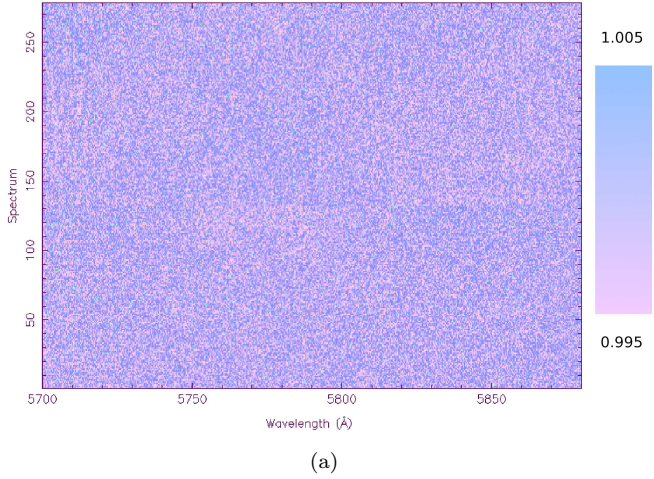}
        }\\%
        \subfigure[]{%
            \label{fig:b}
            \includegraphics[scale=0.31,trim=5mm 0mm 0mm 10mm,clip=true]{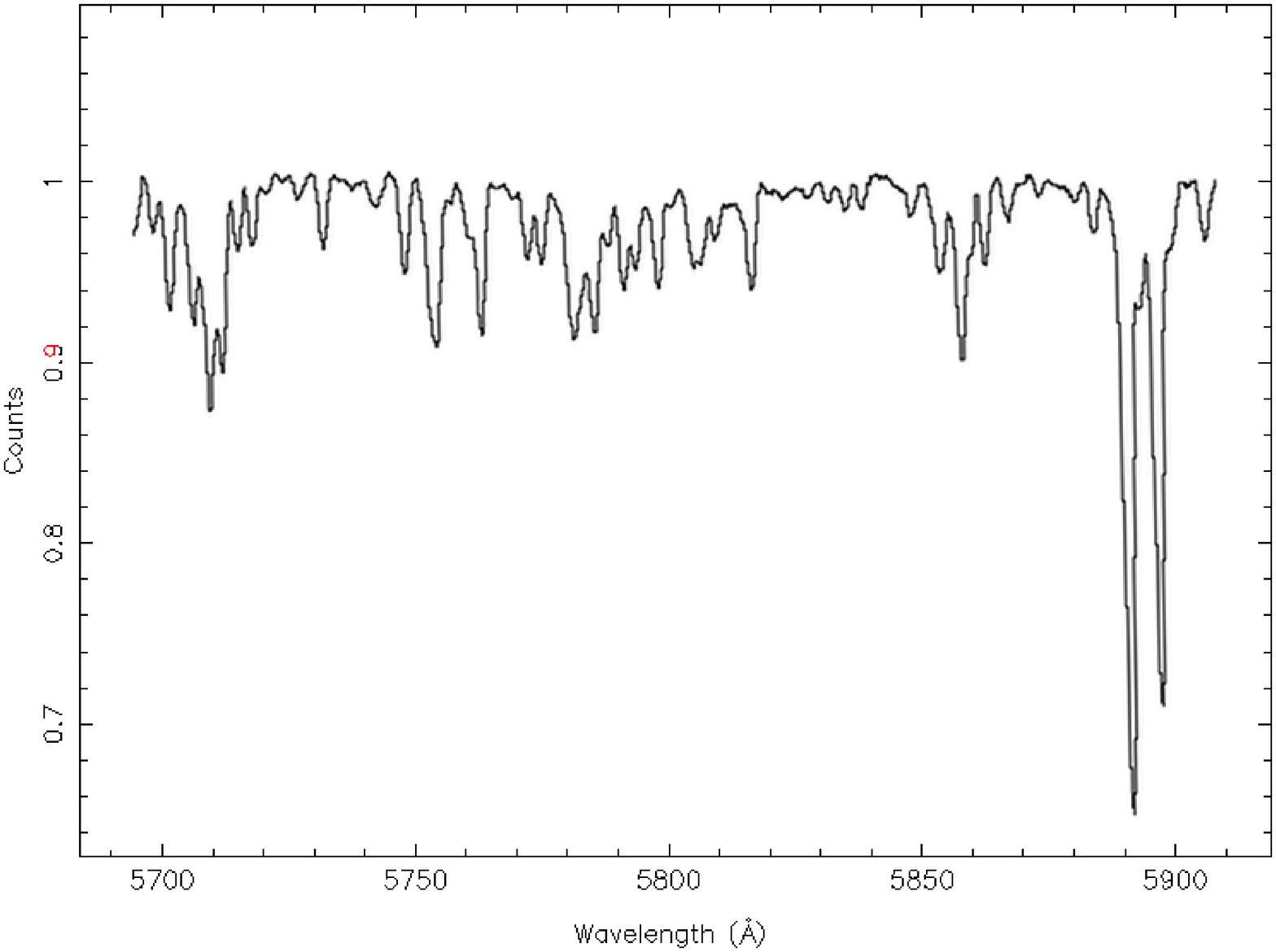}
        }\\ %
        \caption[WASP-12 spectra]{(a) Trail of the residuals after dividing the normalised spectra 
        by the average out-of-transit WASP-12 spectrum. Very little variation appears in the residuals, 
       as indicated by the colour bar. (b) The average WASP-12 out-of-transit spectra. Note that no
        systematics appear in the trailed spectra, indicating that the normalisation has been performed
        correctly, and does not introduce additional sources of error.}
\label{fig:wasp122spectra}
\end{figure}

A number of approaches were taken for comparing the Na and 
continuum windows. The analysis was carried out on spectra 
before normalisation -- in order to search for systematics -- and 
post-normalisation, after the correction for systematics had been 
made. The difference between the spectra post-normalisation in 
both cases (i.e. with and without correction for airmass) is negligible,
indicating that the normalisation does indeed remove any trends which
would adversely affect the search for additional Na absorption. Despite 
this, we opted to use the spectra which had the airmass term removed 
so that any subtle effects will have been removed prior to normalisation. 
The final step before searching the Na region was to ensure that the 
normalisation of the spectra were carried out successfully. Since the 
WASP-12 data were unaffected by variable cloud coverage, a simple 
2-term polynomial could be used in order to normalise the spectra. 
In order to ensure that the normalisation process did not introduce any 
systematics, a thorough investigation into the normalised continuum 
region was carried out. This was done by examining the residuals 
present in the continuum after dividing the normalised spectra by the 
average out of transit spectrum. These residuals were explored to look 
for regions which showed evidence of systematic variation -- something 
which may have been introduced by the normalisation process. After 
performing this analysis, no such systematics were found, indicating 
that the normalised spectra could now be used to search for a Na 
absorption signature. Figure \ref{fig:wasp122spectra} shows the 
continuum region residuals, in addition to the average out-of-transit 
spectra used to obtain these residuals. Note that the trail of the residuals 
appears featureless, indicating that the normalisation process has not 
introduced any additional systematics.

\begin{figure}
\includegraphics[scale=0.21,trim=0mm 0mm 5mm 0mm,clip=true]{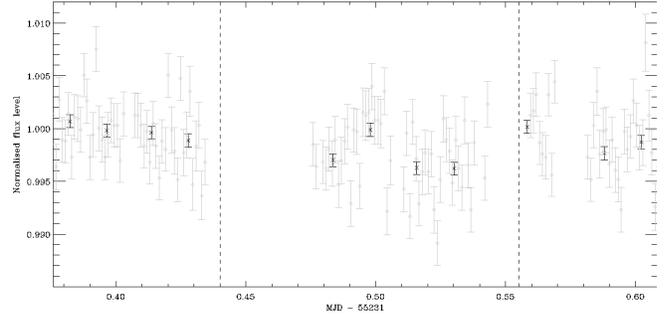}
\caption[Na divided by continuum -- binned]{The Na region divided by the continuum for the
normalised spectra over the first night of observation. The black points are the data after binning, the 
grey points are the data prior to binning. The 
dashed lines indicate the times when the planet is in transit.}
\label{fig:nabinned}
\end{figure}

\begin{figure}
\begin{center}
\subfigure[]{%
            \label{fig:a}
            \includegraphics[scale=0.215,trim=0mm 0mm 0mm 0mm,clip=true]{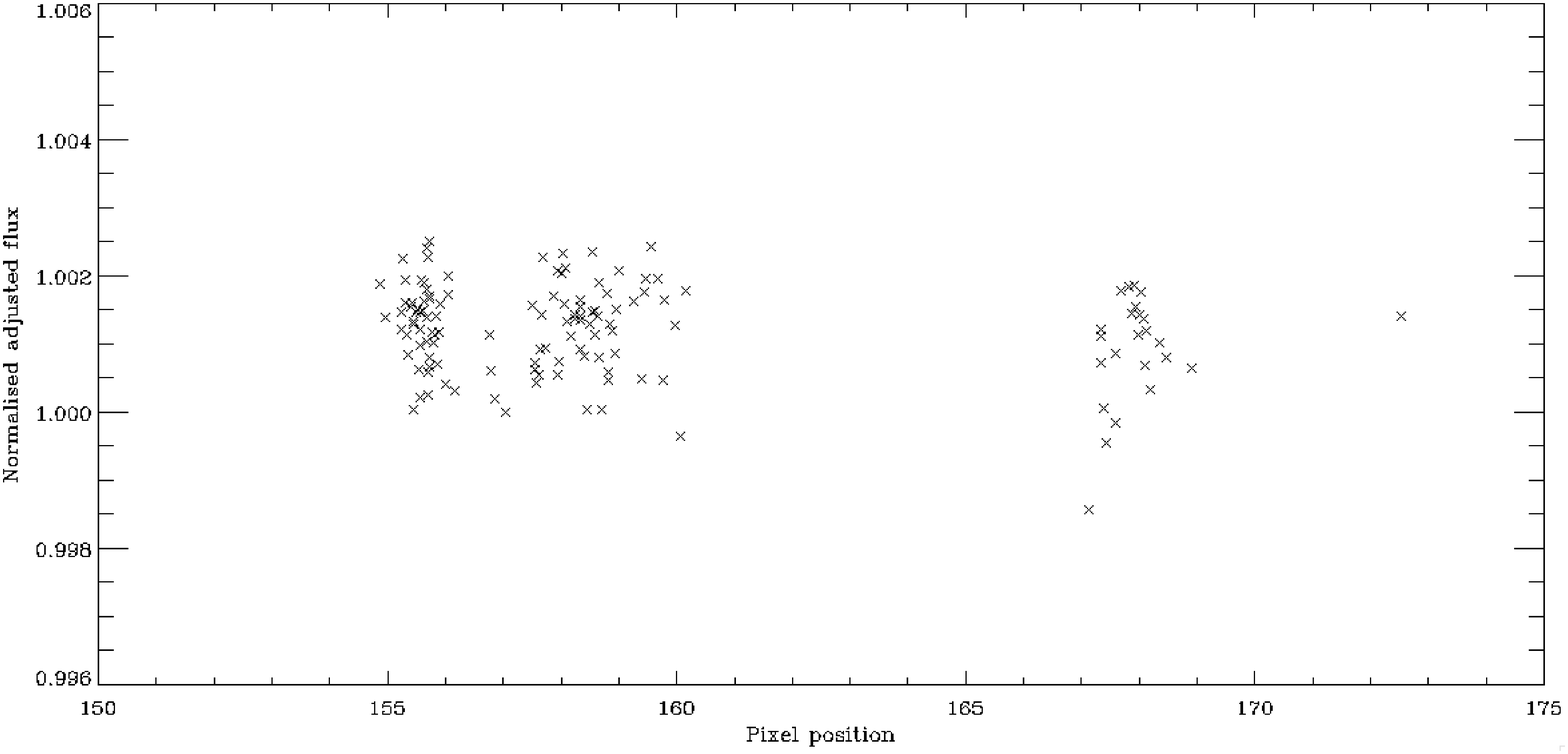}
        }%
        \subfigure[]{%
            \label{fig:b}
            \includegraphics[scale=0.215,trim=0mm 0mm 0mm 0mm,clip=true]{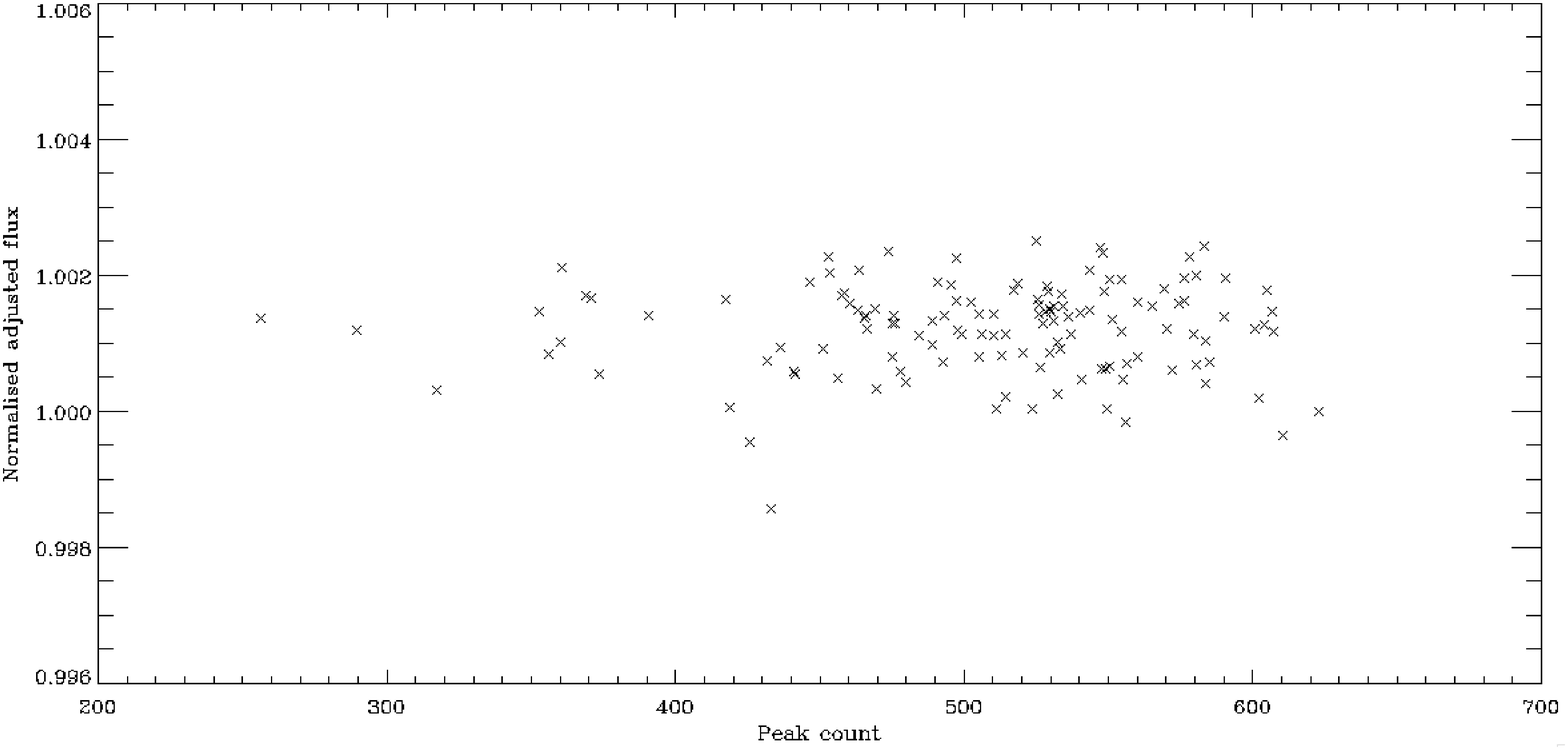}
        }\\ %
        \subfigure[]{%
            \label{fig:c}
            \includegraphics[scale=0.215,trim=0mm 0mm 0mm 0mm,clip=true]{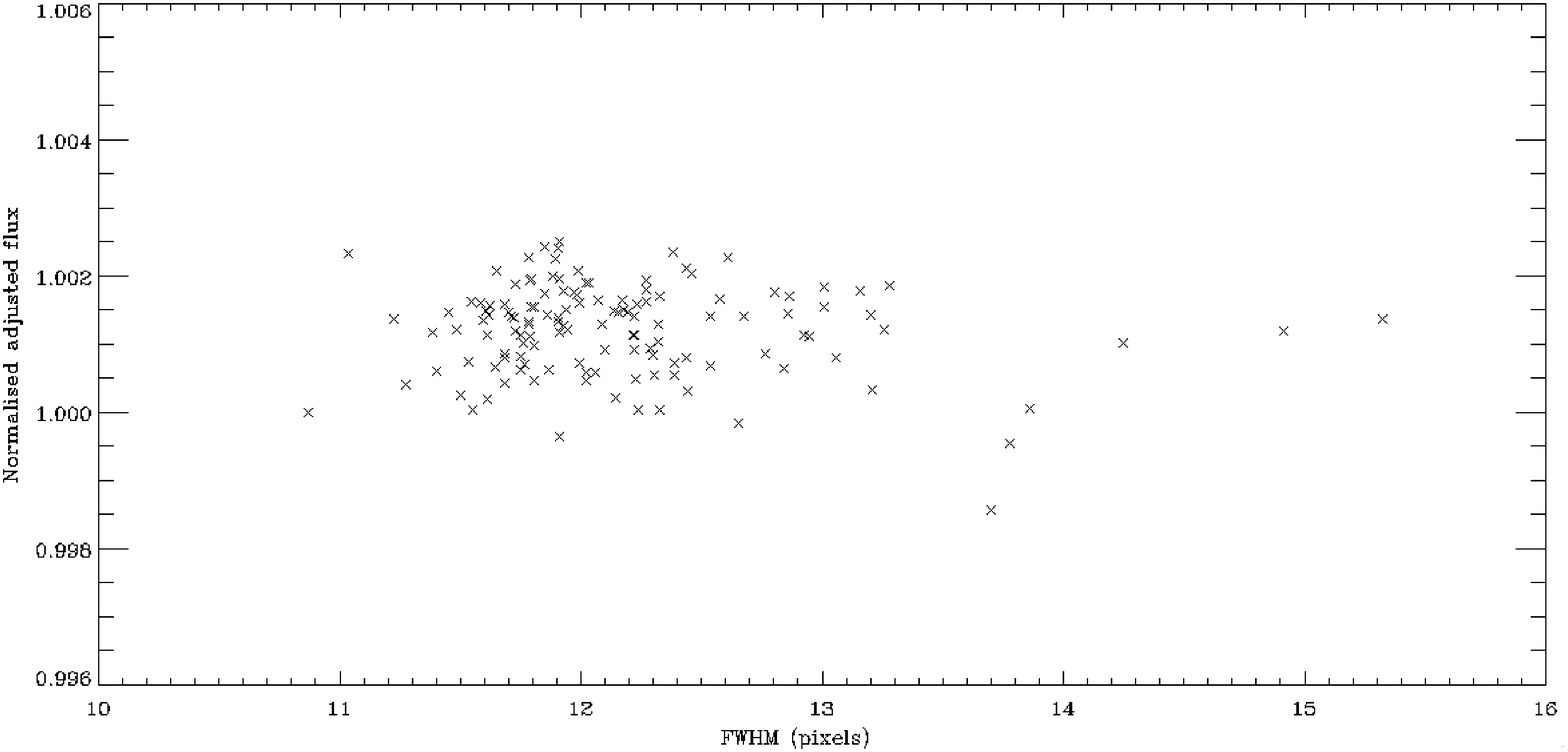}
        }%
        \subfigure[]{%
            \label{fig:d}
            \includegraphics[scale=0.215,trim=0mm 0mm 0mm 0mm,clip=true]{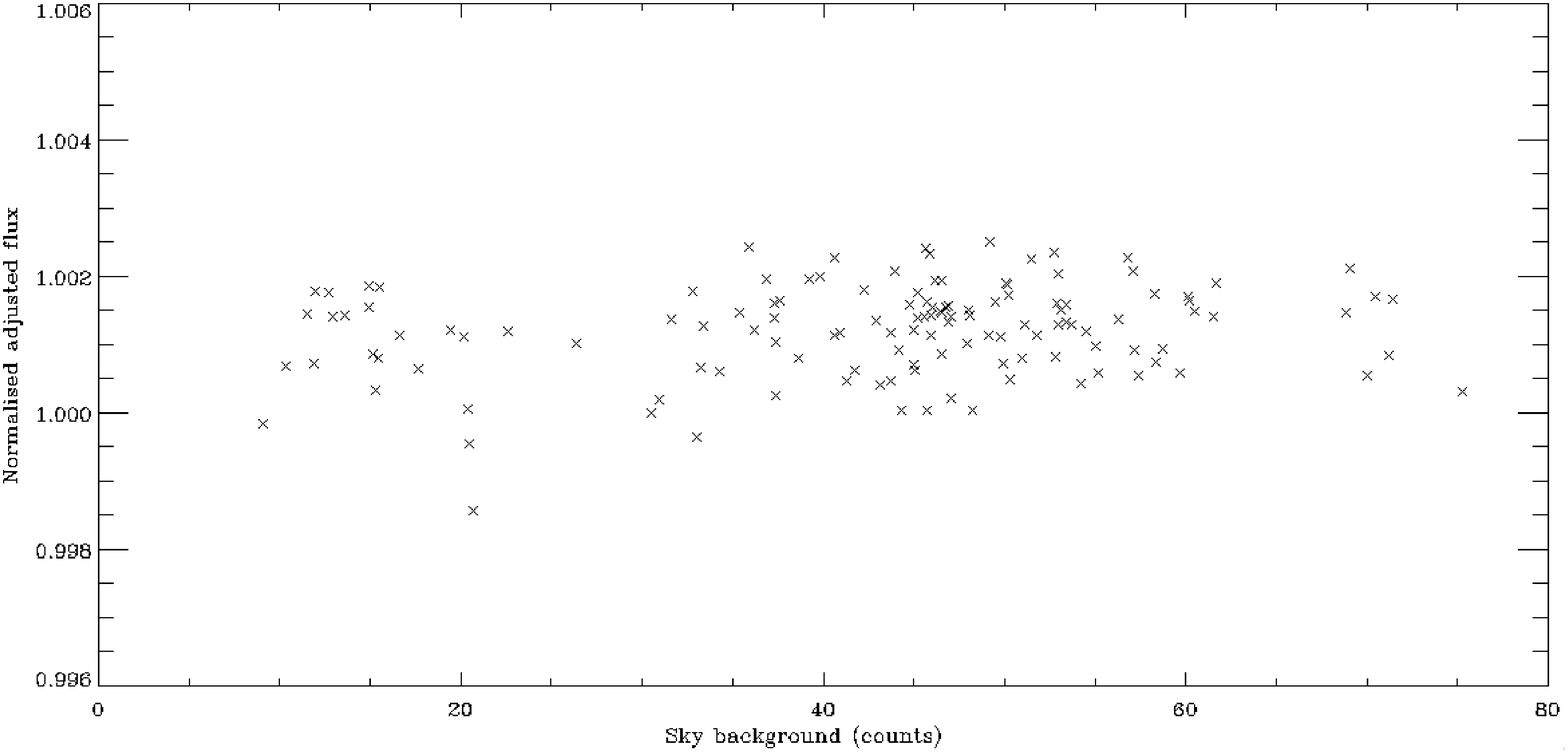}
        }\\%
        \caption[WASP-12 flux correlation analysis]{Night 1 correlations, normalised, corrected flux vs. (a) position. (b) continuum flux. (c) FWHM.
        (d) sky background. Note how there appear to be no correlations post normalisation, given
        the variability of the parameters in Figure \ref{fig:wasp12}.}
\label{fig:wasp12light2}
\end{center}
\end{figure}

The data were then analysed using the Na mask, and the 
continuum region mask (5720 - 5860\AA) in order to generate two sets of normalised 
fluxes -- the Na doublet, and the blueward comparison continuum 
region. A number of different continuum masks were trialled, including masks of the same
width and separation of the Na doublet region, in addition to masks spanning both sides of the
Na feature. The mask we chose to use from this point on in the analysis
gave a large enough coverage so as to ensure a sufficient comparison, but was also 
positioned sufficiently far from the Na region so as to avoid
potential interference.
Figure \ref{fig:nabinned} shows the ratio of the light under 
the Na mask to the continuum region. During transit, this ratio 
decreases, indicating that the Na lines are deepening. As the planet 
moves out-of-transit, the flux ratio increases back to its pre-ingress value. 
The second binned point in transit which appears at a similar level to the 
out-of-transit points occurs at the same time as a drop in the flux at mid-transit 
(see Figure \ref{fig:wasp12}), 
thought to be due to transparency
variations. The correlations of Na/continuum flux ratio with 
position, continuum flux, seeing and sky background were thoroughly
investigated, and no trends could be found, either before or after
normalisation (see Figure \ref{fig:wasp12light2}). Again, since this is
based on only one night's data, low-level correlations may become 
apparent as more observations are made.

\begin{figure}
\includegraphics[scale=0.21,trim=0mm 0mm 0mm 0mm,clip=true]{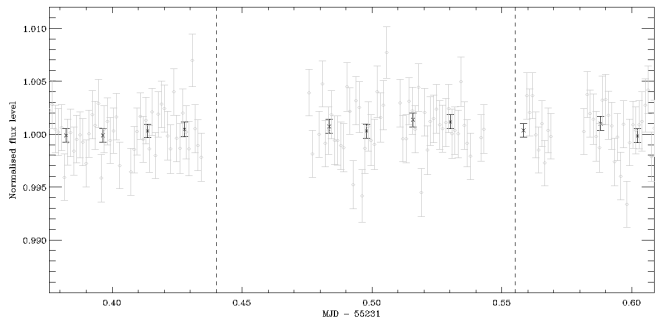}
\caption[Masked region divided by continuum -- binned]{A featureless region of the 
continuum from 5800 - 5810{\AA} has been analysed as opposed to the Na region. Again,
the black points are the binned data and the grey points are the unbinned. Note that in this
case, the points are almost constant, without the deepening seen in Figure \ref{fig:nabinned}.
This analysis was performed over a number of different regions in the spectra, and similarly,
no deepening could be found (see Figure \ref{fig:diff}).}
\label{fig:contbinned}
\end{figure}

In addition, we also performed a check using the same Na mask on a number of featureless
regions in the continuum. Since there should be no variation in the depth of the normalised 
continuum features, this analysis gives an excellent check to determine whether this deepening is
unique to the Na region. Despite using the same mask widths, and using a large range
of the continuum to perform this analysis, the decrease in flux ratio we see in the Na region
does not occur anywhere else in the continuum. Figure \ref{fig:contbinned} is a similar plot to Figure \ref{fig:nabinned},
but rather than centering on the Na doublet, the Na mask has been set to a featureless
region of the continuum (5800 - 5810{\AA}). We also tested the window using regions away from the
comparison region, on both sides of the Na doublet to ensure there was no variation in the ratio.
Upon testing these regions, the ratio showed no such change as in Figure \ref{fig:nabinned}.

\begin{figure*}
\centering
\includegraphics[width=\hsize,trim=2mm 0mm 0mm 0mm,clip=true]{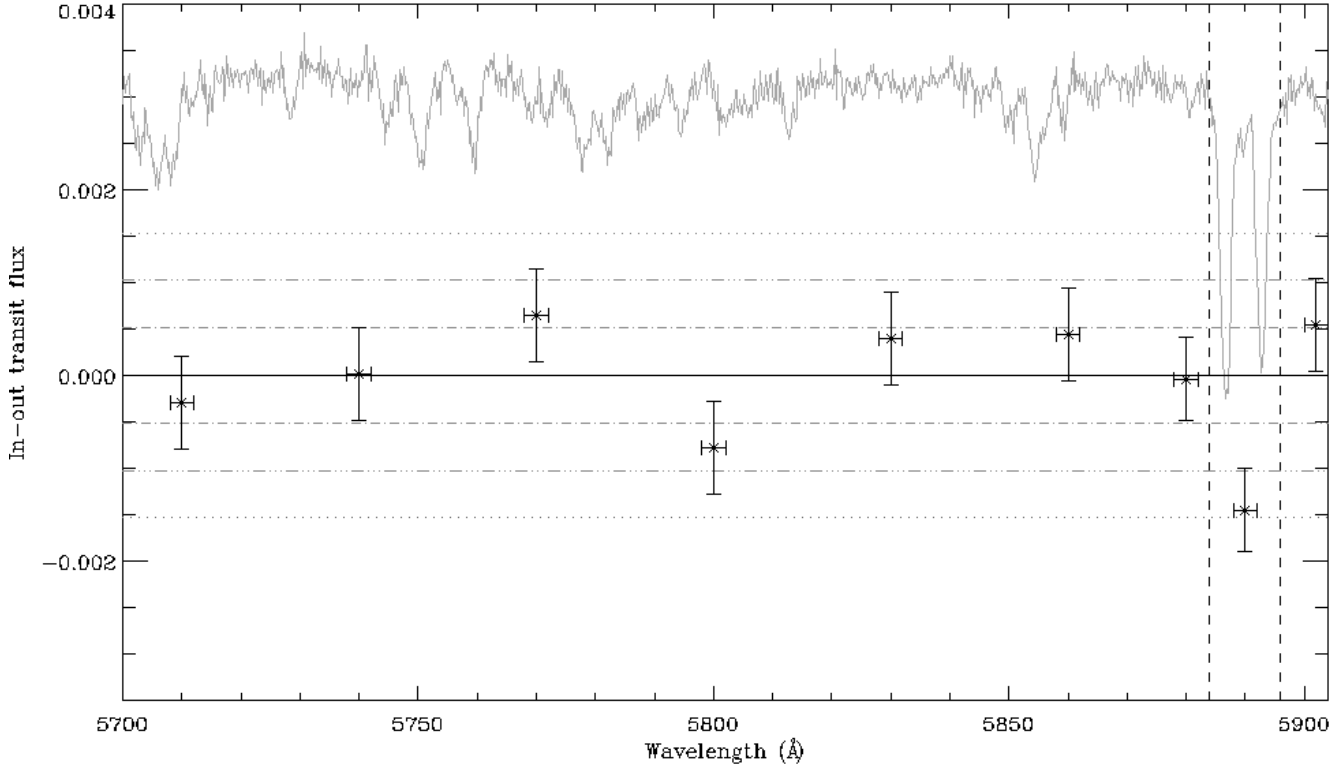}
\caption[Difference between in- and out-of-transit flux ratios over the spectra]{The difference
in flux ratio (in-out of transit flux) 
for a series of windows in the comparison region for the Na feature (highlighted by
the vertical dashed lines). The horizontal error bars represent the wavelength coverage of the 
mask in the spectrum analysis.
The dot-dash, dot-dot-dash and dotted horizontal lines are the 
1$\sigma$, 2-$\sigma$ and 3-$\sigma$ levels, respectively.
The points have been overlaid on an example WASP-12 spectra for an indication
of where the windows were placed. Note how the Na region is the only one which shows a significant
difference between the in- and out-of-transit depths, lying in the 3$\sigma$ region. The comparison
windows have been taken at 30\AA\ intervals along the continuum.}
\label{fig:diff}
\end{figure*}

The result of these tests are almost identical no matter where in the 
continuum the Na mask is placed. In Figure \ref{fig:diff}, we show the difference between the in- and out-of-transit flux ratios for the
windows that were chosen along the continuum, as well as when the window covers the Na region. 

We can therefore state that the increased absorption during transit is unique to the 
Na doublet for this data set. 
Whether this is an effect of the atmosphere of WASP-12b, or a still-unknown 
systematic only affecting the Na region (such as 
an issue caused by linearity) is still unknown, and will require further observations
to fully confirm or deny any claim to the detection of the atmosphere. 
For example, changes in the stellar spectrum due to star spots
could cause interference in the Na line depth. This could have the effect of either mimicking
a Na detection from the exoplanet atmosphere, or interfering with the level of Na absorption
during transit. 
The umbral region of a starspot on a G-type star covering a stellar surface area of 20$\times10$$^{-6}$$A_{1/2\odot}$
has been shown to dilute the Na II
line depth relative to the continuum region by 0.033$\pm$0.05\% (\citealt{fay72}). This 
indicates that a particularly high level of stellar activity during observations could mimic 
the level of Na absorption
found during previous transmission spectroscopy studies (e.g. 
this should be compared to the Na transit signature found for HD209458b, 
0.135$\pm$0.017\%, \citealt{snellen08}). During previous observations however, WASP-12 has 
displayed low-levels of stellar activity (\citealt{haswell12}, \citealt{sing13}), indicating that any Na II deepening at the levels
comparable to those found by \cite{snellen08} are unlikely to be solely due to
sunspots.
In addition, Na II deepening
caused by stellar variability would not only affect the spectra during transit, and hence monitoring
large portions of out-of-transit data would also provide a method to characterise short-period
changes in the Na lines due to this effect.

One could also argue that the deepening around the Na doublet could be caused by
tellurics, since our search is in the wavelength region where these lines start to become an issue 
($\gtrapprox$5885{\AA}). However, when we move our masks to continuum regions which should also 
be affected by tellurics -- 5880-5887{\AA} and above 5900{\AA}, no additional deepening was detected
when performing our flux ratio analysis.
Placing masks in this region also has the additional advantage of checking for instrumental effects 
in the part of the spectrum near the edge of the CCD. We considered carrying out a similar telluric analysis to that performed by \cite{snellen08}, where a synthetic
telluric spectrum was constructed, matched to the resolution of the data, and monitored by
comparison with a telluric standard. However, the slit width we used for our arc line acquisition
made lines `top-hat' shaped, meaning we are unable to accurately model the instrumental broadening 
profile of our spectra. 

Another effect one must consider is that of limb-darkening. If the planet transits an optically thick
part of the star, the continuum will deepen as compared to the Na line. Previous investigations
have revealed that the effect of limb-darkening is minimal (\citealt{charbonneau02}, 
\citealt{snellen08}) when investigating the continuum surrounding the Na region.
Our observations were also set-up to specifically avoid an unknown source of absorption
redward of the Na doublet (as reported by \citealt{sing08}), which would have been significantly
worse than the limb-darkening signal. \cite{redfield08} find the impact minimal, with
excess limb darkening absorption contributing +1.46$\times$$10^{-5}$ to their Na signal
of (-67.2$\pm$20.7)$\times$$10^{-5}$. The limb darkening for our measurements should fall
well within the error on any detection of the Na absorption at the level of $\sim$0.005\%.
Finally, the contribution of the companion objects to WASP-12 to the Na region must be considered.
As mentioned in Section \ref{sec:intro}, WASP-12 is part of a triple system, meaning the companions
to the WASP-12b host star may impact on the Na measurements. \cite{bechter14} find the
two objects to be M3V dwarfs, with a separation of 84.3$\pm$0.6 mas. 
In order to investigate the specific effect of the M-dwarf pair on both
the continuum region and the Na lines, we estimated the dilution factor in the wavelength 
range 5700--5920\AA.
Assuming all three objects are at the same distance away (as found by \citealt{bechter14}),
we obtained synthetic spectra for a G0 star and an M3V dwarf using the PHOENIX NextGen2 
model atmosphere grid (\citealt{hauschildt99}). We matched the temperature, metallicity and 
log g values to the nearest observed values taking a cooler temperature for WASP-12 
(since NextGen2
increases in 200k increments) so as not to underestimate the contribution. By comparing the
synthetic spectra in this region, we estimate that the fractional contribution in the Na region by the M-dwarfs
is negligible and in the nearby continuum region is 1.7\% for each M-dwarf in the system. 
Obtaining more observations of this system would allow for an opportunity to monitor
the continuum region to see if the contribution is stable or changes over time. Any future
defocussed transmission spectroscopy observations which aim to detect planetary
atmospheres around multiple-star systems should be mindful of this effect and attempt to
estimate the fractional contribution of any companions.

If we are to assume that the deepening of the Na lines for the first night is due to the
atmosphere of WASP-12b, we can obtain the corresponding sodium absorption.
We obtain a value of 0.12$\pm$0.03\%, a value which is very similar to the narrow
absorption band of \citealt{snellen08} for HD209458b (0.135$\pm$0.017\%). 
In order to check that this result is physically consistent with a real atmosphere detection, 
the data were analysed using wider masks. Increasing the size of the
wavelength region being analysed has the effect of decreasing the effective Na cross
section, resulting in smaller absorption signals. This has been the case with most
Na exoplanetary atmosphere detections (\citealt{charbonneau02}, \citealt{snellen08}, 
\citealt{redfield08}, \citealt{sing08}). Increasing the size of the wavelength mask surrounding
the Na region also has the effect of analysing more of the redward and blueward regions,
providing a further investigation into the surrounding Na features. 
The results of this are shown in Table \ref{tab:na}.

\begin{table}
\centering
\caption[Table]{Results of the analysis using masks
of increasing size. Note that the error here has been 
adjusted to include uncertainty arising from the systematic
spike present mid-transit.}
\begin{tabular}{ccc}
Mask Width & Absorption signal & Error \\
\AA & \% & \% \\ 
\hline
2 & 0.12 & 0.03[+0.03]\\
4 & 0.04 & 0.022[+0.03]\\
10 & 0.01 & 0.012[+0.03]\\
\label{tab:na}
\end{tabular}
\end{table}

\begin{figure}
\includegraphics[scale=0.26,trim=0mm 0mm 0mm 0mm,clip=true]{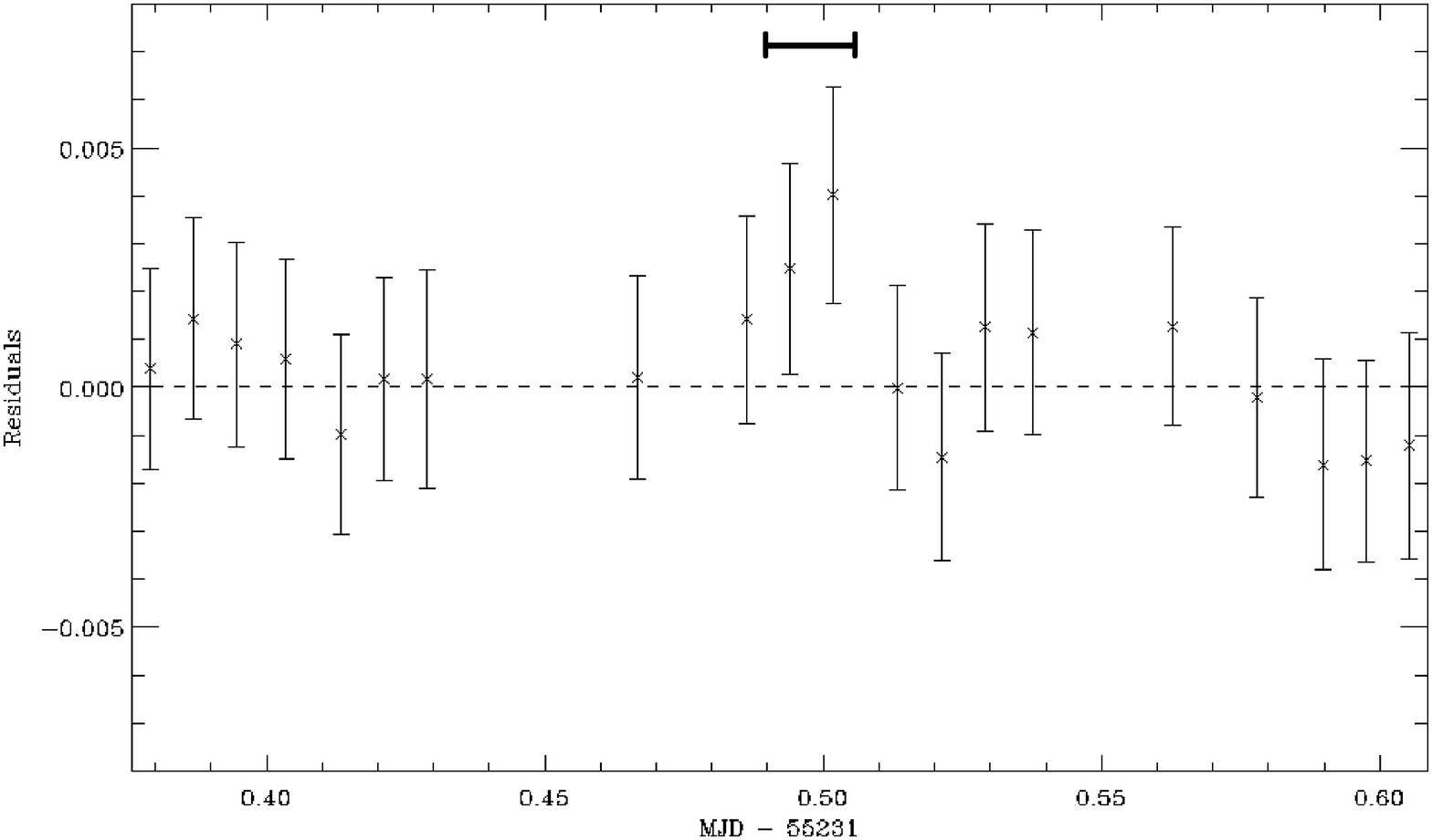}
\caption[Bin test]{The residuals returned from the light curve analysis using a binning factor of 6. 
This bin factor was determined using the method outlined in \cite{pont06}. Note the points which
rise mid-transit, causing the absorption depth to lower. We concentrated or analysis on these
series of points (highlighted above with the bar range) to determine the potential error they introduced.}
\label{fig:bintest}
\end{figure}

The values we return for the 4 \& 10\AA\ masks do indeed follow with what
we expect, with the absorption signal decreasing for the wider masks. 
The 4\AA\ mask shows a lower absorption depth (0.04$\pm$0.022) 
and the 10\AA\ mask shows no additional absorption (0.01$\pm$0.012).
However, whilst these results do show
what one would expect, it is important to note that since this is still based on a single night's data,
these absorption depths may be liable to systematics we are unable to account for,
and therefore need further observations to check whether this is a real physical effect
due to the atmosphere.
We note that our figure also
includes the rise in the ratio during the transit, thought to be due to the brightening
sky background. This means our figure of 0.12\% is diluted by this systematic effect.
We also note that 0.03\% is our formal error using the error from the data points, 
and will increase due to the systematic feature mid-transit. In order to characterise the
effect of this systematic, the model light curve used to calculate our absorption depth 
was subtracted from the data. These residuals were binned (in the manner prescribed by
\citealt{pont06}), and the mid-transit systematic feature determined from this. 
Figure \ref{fig:bintest}
shows the result of this analysis using a binning factor of 6. As can be seen from this figure,
the residuals at the mid-transit point are the only ones to show significant positive deviation 
from 0, indicating the region most affected by this unknown source of noise. Isolating this
series of data and removing it from the absorption analysis gives a new absorption depth
of 0.15$\pm$0.05\%, a difference of 0.03\% from the unaltered case. We have
indicated the effect of this systematic on our Na absorption depth by indicating the 
additional systematic error alongside the statistical error i.e. 0.12$\pm$0.03[+0.03]\%.
This is a significant 
effect, and indicates the influence that systematics can have on a single data set. Whilst this
is still a physically reasonable absorption signal, removing points has increased the error.
This type of analysis should be employed during future observations to check if the data are
similarly effected by such systematics at any point during the transit. 

Since we expect the Na absorption of WASP-12b to be slightly higher than that of
HD209458b (due to the large scale height we are probing in addition to the extreme
distortion and mass loss of the atmosphere), this seems a
reasonable figure, given the systematic occurring mid-transit.
In order to place further statistical significance on the Na detection, we opted to take a
Monte Carlo approach. We selected a number of random out-of-transit spectra from our Na region, 
and used this in place of our in-transit selection. Carrying this out over 25,000 iterations gave
excellent estimations of the false alarm statistics present in the data. This 
gave the 68\%, 95\% and 99.7\% bounds for our data, which
are shown by the dotted and dot-dashed lines in Figure \ref{fig:diff}. As can be seen, the Na point does
lie within the 99.7\% population, indicating that the deepening of the lines is unlikely
to be due to random error. 

\section{Defocussed Transmission Spectroscopy as an Observing Technique}\label{DTS}

An important aspect of the research carried out during these observations was
investigating the technique of defocussed transmission spectroscopy. Whilst
the observations themselves were marred with poor weather conditions and
technical issues, a number of issues were discovered which will enable future
runs to be carried out much more efficiently. In addition to this, the initial exploration
of the data has lead to a potential detection of Na in the atmosphere of WASP-12b, 
indicating that the technique is feasible
on telescopes the size of the WHT. Features such as the rise in flux ratio mid-transit
also provide areas which could be investigated during future observations. Is this a
real feature, or simply an unknown systematic? Our statistical analysis does indicate
that it has a significant effect on the absorption depth
In regard to the technique of Defocussed Transmission Spectroscopy, we have demonstrated
that there does exist extensive justification for the refinement, development and continued
use of the technique based on our data. Further observations will be able 
to improve upon both the technique and data analysis to make this a powerful method
of observation for characterising exoplanetary atmospheres.

\section{Future Observations}\label{sec:future}

Our pilot study of defocussed transmission spectroscopy revealed a number of improvements
to the technique which could be easily applied to future observations. This section
details some additional steps one should take during data acquisition and analysis
to further improve the technique.

\begin{figure}
\includegraphics[scale=0.255,trim=0mm 0mm 0mm 0mm,clip=true]{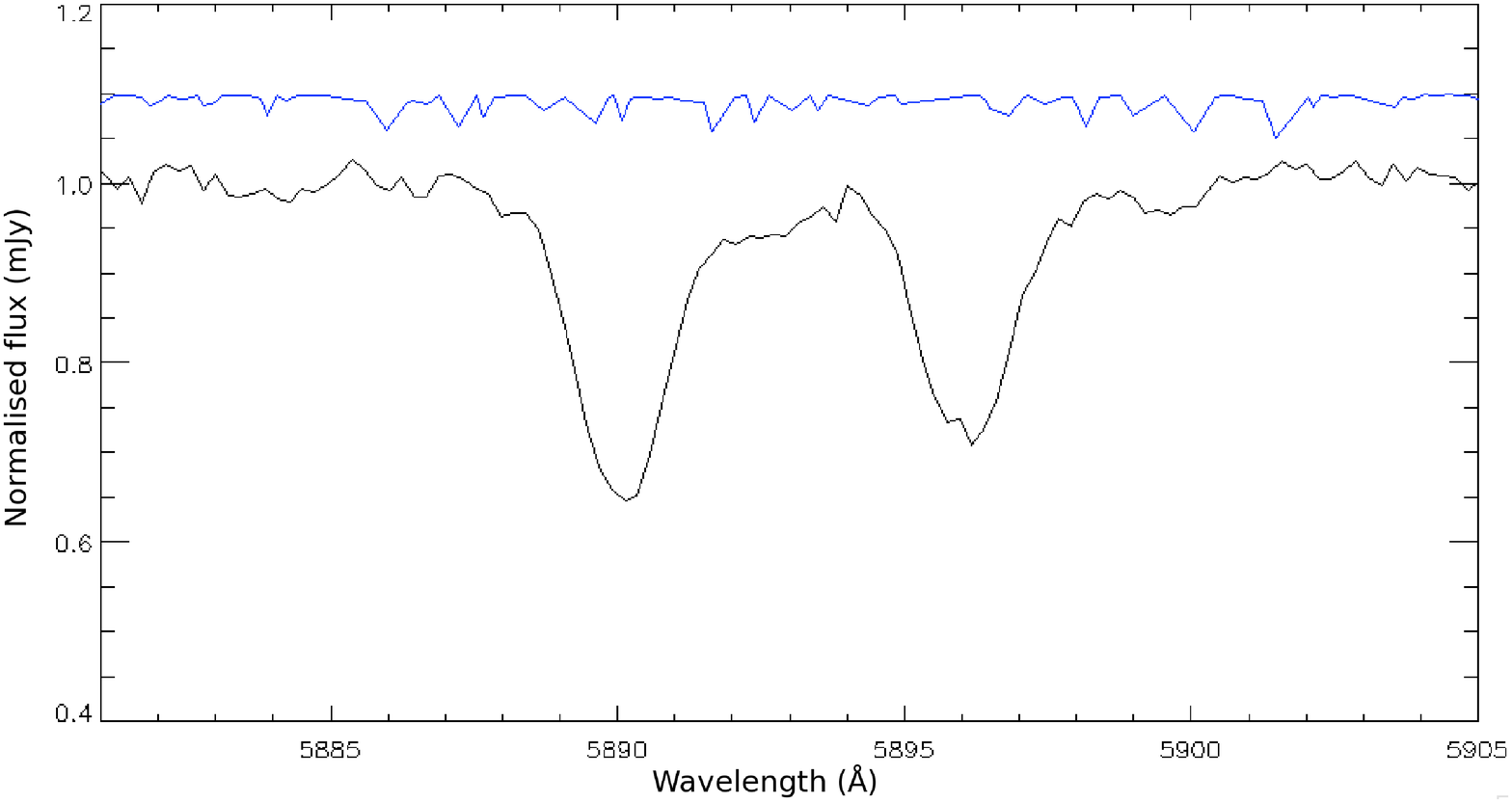}
\caption[Telluric monitoring]{The region where we expect telluric contamination to potentially
deepen the lines. The top blue line is a synthetic telluric spectrum created from the line lists
of \cite{lundstrom91}. The bottom plot is a WASP-12 spectrum showing the Na doublet. Note how
the telluric features will effect the both the Na lines themselves and the continuum surrounding 
these lines. We note here that due to the resolution of the synthetic spectra, the lines are only representative of their position, not necessarily their depth.}
\label{fig:telluricspec}
\end{figure}

In order to fully characterise any potential contamination due to tellurics, it is desirable to follow the
method prescribed by \cite{snellen08}. In order to model the tellurics using a synthetic telluric spectrum, the line-list needs to be convolved with the instrumental broadening profile. While this could be obtained from the arc lines, the arc frames
during our observations were taken with a wide slit, meaning the arc lines appear as `top-hats'
and hence the profile could not be accurately obtained.
Figure \ref{fig:telluricspec} shows how tellurics would be modelled if it were possible to
convolve the line-list with the instrumental broadening profile for the data. 
The top (blue) spectrum is a synthetic telluric spectrum from \cite{lundstrom91},
along with a WASP-12 spectrum centred on the Na doublet.

It would also be desirable to obtain a data set of a bright target (for example, HD189733) under photometric conditions, with ample out-of-transit data. This would allow for the
systematics of the technique to be further investigated and characterised. Such a dataset could
allow for additional systematics (such as instrumental effects) to be modelled by injecting
a fake transit into the spectra, and attempting to recover the signal. Such techniques would provide
a useful investigation into additional methods of attempting to detect absorption due to an
exoplanet atmosphere.

When observing the transit, future observations should aim to get as much
out of transit data as possible in order for any potential sodium detection to have a large number
of continuum regions to compare to. For our initial observing run, we were limited to two half nights, due to the position of the object on the sky. Observations of the full transit and ample out-of-transit spectra would allow
for any trends in the Na/continuum flux ratio to be monitored both prior to ingress and post-egress.
It would also be useful to obtain a data set of in-focus transmission spectroscopy observations using the same instrumental set-up. This 
would provide a direct comparison between the techniques, and would give an indication as to the
improvement defocussing has on observations. In addition, better characterisation of the linearity of the CCD would
allow for an explicit separation of airmass correction and linearity correction during data
analysis.

The C$_{1}$/C$_{2}$ correction was applied by removing
the trend in the comparison region of the spectrum, extrapolating this to the whole
spectrum prior to normalisation. Doing this does ensure a better comparison region,
however, extrapolating towards the Na region is not ideal, as there may be some residual
curvature in the spectrum which this method does not take into account. In order to provide
a thorough check of all parts of the spectrum, future observations should aim to obtain coverage
redwards of the Na doublet which can be compared with the blueward spectrum to fully
investigate any curvature in the spectrum prior to correction. Whilst this region may be 
prone to an unknown source of absorption, it does not have to be used in the Na/continuum
analysis, and provides an additional investigation of the systematics. 

\section{Conclusions}\label{conc}

The technique of defocussed transmission spectroscopy has the capacity to allow
for the characterisation of exoplanet atmospheres of fainter targets, providing
observational constraints to atmospheric models. Any tentative detection 
of Na in the atmosphere of WASP-12b is a significant result,
providing a potential insight into this highly irradiated, highly inflated hot-Jupiter.
This detection would also provide a direct observational test of the
pM/pL classification system. Not only should Na
not be detectable, due to TiO/VO swamping Na at these wavelengths, but the level we
detect it at indicates that WASP-12b should fall into the pL category. This runs directly
contrary to the predictions made by \cite{fortney08} for such an irradiated system, 
and provides evidence that the classification scheme may need reappraisal.
Recent observations by \cite{sing13} also indicate the need for reappraisal due to the 
lack of TiO in the atmosphere of WASP-12b.
In addition, constraints on the P-T profile of the atmospheres of hot-Jupiters can be obtained from
measurements of Na, giving further evidence for or against the predictions of atmospheric
models. Recent HST Wide Field Camera 3 (WFC3) WASP-19b transmission spectroscopy
observations find a lack of TiO absorption features, and
a signature of H$_{2}$O in the atmosphere (\citealt{huitson13}). WASP-19b, 
like WASP-12b, is predicted to lie in the pM region of the classification scheme of \cite{fortney08} and should,
therefore, be dominated by TiO absorption. The claim of \cite{huitson13} that TiO is reduced in
the atmosphere of WASP-19b runs contrary to the predictions of the pM/pL classification
scheme, much like the WASP-12b Na detection presented in this paper.
Both of these results provide an indication that the classification system 
should be reconsidered. Further transmission spectroscopy measurements will allow for more 
observational points to revise this, or indeed any, classification scheme.

From the extensive analysis of systematics and correlations for the data, a number of 
preliminary areas for investigation as to how to
characterise and monitor systematic variations during future observations have been identified.
The effects of both short- and long-term variations during observing have been noted, in addition
to investigating correlations between the flux and available parameters. However, due to poor weather conditions and the low-resolution
of the data, further observations must be made in order to obtain a better understanding of how
systematics such as telluric contamination and linearity of the detector affect the fully reduced
spectra. 
We have eliminated a number of potential causes which could mimic the Na absorption
effect during transit, and investigated other potential causes, such as deepening due to
telluric contamination and linearity. 
Whilst there does not appear to be any evidence of these systematics, 
more data 
at a higher
resolution is desirable in order to fully eliminate these. 

Overall, the detection of Na at the level presented in this analysis is potentially a very significant  
result. Based on a single night's data, there does appear to be a possible detection of additional
Na absorption due to the atmosphere of WASP-12b. However, more observations are
required before such a detection is confirmed. Nonetheless, the technique of defocussed
transmission spectroscopy does appear to be a viable method for the detection
and characterisation of exoplanet atmospheres. Further use of the technique will allow
for it to be refined to make direct detections from ground-based platforms.

\section*{\sc Acknowledgements}

The authors gratefully acknowledge the use of Tom Marsh's software packages
\textsc{molly} and \textsc{pamela}.
JB was funded by the Northern Ireland Department of Employment and learning.
CAW acknowledges support from STFC grant ST/L000709/1.
PRG is supported by a Ram\'on y Cajal fellowship(RYC--2010--05762), and
also acknowledges support provided by the Spanish MINECO under grant 
AYA2012--38700.
The WHT is operated on the island of La Palma by the Isaac Newton Group in the Spanish Observatorio del Roque de los Muchachos of the Instituto de Astrofsica de Canarias.
We would especially like to thank the referee for numerous helpful comments
after initial submission of this work.

\bibliographystyle{mn2e}
\bibliography{abbrev,refs}

\nopagebreak 

\onecolumn

\begin{table}
\centering
\caption[Observations]{Observations of the three candidates in our pilot study.
Note that switching the observing mode from
`Fast' to `Slow' was due to an issue with the CCD controller.}
\begin{tabular}{lcccccccc}
Target & Transit & Slit width & Flat-fields & Exposures & MJD start & Exp time (s) & Read mode & Observing Conditions \\ 
\hline
WASP-12 & In-transit & 2.48" & 1055 & 53 & 55231.476 & 101 & Fast & Some light cloud \\
 & Out-transit &  &  & 79 & 55231.376 & 101 & Fast & Some light cloud, mostly clear\\
 & In-transit & 2.48" & 1004 & 48 & 55232.526 & 101 & Fast & Some light cloud, mostly clear \\
 & Out-transit &  &  & 98 &  55232.396 & 101 & Fast  & Mostly clear\\
HD209458 & In-transit & 2.48" & 2226 & 96 & 55445.452 & 51 & Slow & Light cloud \\
 & Out-transit &  &  & 202  & 55445.373 & 51 & Slow & Cloudy\\
HD189733 & In-transit & 2.48" & 1205 & 239 & 55446.367 & 51 & Slow & Variable cloud\\
 & Out-transit &  &  & 52 & 55446.359 & 51 & Slow & Light cloud\\
\label{tab:tscand}
\end{tabular}
\end{table}

\end{document}